\documentclass[a4paper]{article}
\pdfoutput=1
\usepackage{jheppub}
\usepackage[T1]{fontenc}
\usepackage{graphicx}
\usepackage{amsmath}
\usepackage{xspace}
\usepackage[numbers,sort&compress]{natbib}
\usepackage{subfig}
\usepackage{pstricks}
\usepackage{array}

\newcommand{\del}[1]{}

\definecolor{darkgreen}{rgb}{0,0.5,0}
\definecolor{darkpurple}{rgb}{0,0.5,0.5}
\definecolor{darkblue}{rgb}{0,0,0.7}
\definecolor{darkred}{rgb}{0.5,0,0.0}
\definecolor{darkorange}{rgb}{0.8,0.4,0.0}
\definecolor{green}{rgb}{0.0,0.8,0.4}

\newcommand{\mh}{{m_H}}
\newcommand{\mhsq}{{m^2_H}}
\newcommand{\pt}{p_{\perp}}
\newcommand{\mb}{{m_b}}
\newcommand{\mt}{{m_t}}
\newcommand{\as}{\alpha_s} 
\newcommand{\abar}{\bar{\alpha}} 

\newcommand{\MS}{\overline{\rm MS}}
\newcommand{\be}{\begin{equation}}
\newcommand{\ee}{\end{equation}}
\renewcommand{\gtrsim}{\mbox{\raisebox{-0.3ex}{%
\footnotesize $\:\stackrel{>}{\sim}\:$}} }
\renewcommand{\lesssim}{\mbox{\raisebox{-0.3ex}{%
\footnotesize $\:\stackrel{<}{\sim}\:$}} }
\newcommand{\gsim}{\gtrsim}
\newcommand{\lsim}{\lesssim}

\preprint{
  \begin{flushright}
    IPPP/18/27, CERN-TH-2018-089, TTP-18-016
  \end{flushright}
}

\author[a]{Fabrizio Caola}
\author[a]{Jonas M. Lindert}
\author[c]{Kirill Melnikov}
\author[b]{Pier Francesco Monni}
\author[b]{Lorenzo Tancredi}
\author[c,d]{Christopher Wever}

\affiliation[a]{Institute for Particle Physics Phenomenology, 
		Durham University, 
		Durham DH1 3LE, 
		UK}
\affiliation[b]{TH Division, 
                Physics Department, 
                CERN, 
                CH-1211 Geneva 23, 
                Switzerland}

\affiliation[c]{Institute for Theoretical Particle Physics (TTP), 
		KIT, Karlsruhe, Germany}

\affiliation[d]{Institut f\"ur Kernphysik, KIT, 76344 Eggenstein-Leopoldshafen, Germany}

\emailAdd{fabrizio.caola@durjam.ac.uk}
\emailAdd{pier.monni@cern.ch}
\emailAdd{jonas.m.lindert@durham.ac.uk}
\emailAdd{kirill.melnikov@kit.edu}
\emailAdd{lorenzo.tancredi@cern.ch}
\emailAdd{christopher.wever@kit.edu}

\title{Bottom-quark effects in Higgs production at intermediate transverse momentum}

\abstract{ We provide a precise description of the Higgs boson
  transverse momentum distribution including top and bottom quark
  contributions, that is valid for transverse momenta in the range
  $\mb\lesssim\pt\lesssim\mt$, where $\mb$ and $\mt$ are the bottom
  and top quark masses.  This description is based on a combination of
  fixed next-to-leading order (NLO) results with next-to-next-to-leading logarithmic (NNLL) transverse momentum resummation.  We show
  that ambiguities in the resummation procedure for the $b$-quark
  loops are of the same order as the related fixed-order
  uncertainties.  We conclude that the current uncertainty in
  the top-bottom interference contribution to the Higgs transverse
  momentum spectrum is $\mathcal O(20\%)$.

}

\begin{document}

\maketitle
\flushbottom

\section{Introduction}
\label{sec:intro}
Understanding the Higgs particle observed at the LHC requires studies
of its properties that include quantum numbers and couplings
to gauge and matter fields. Current experimental results have
relatively small ${\cal O}(10\%-20\%)$ uncertainties in Higgs couplings
to electroweak gauge bosons and larger ${\cal O}(100\%)$ uncertainties  in Higgs
Yukawa couplings, especially for light
quarks~\cite{Khachatryan:2016vau}.  However, it is quite conceivable
that physics beyond the Standard Model manifests itself in smaller,
few percent, contributions to Higgs couplings.  Thus, facilitating
further improvements in extracting Higgs couplings to gauge bosons and
helping constrain Yukawa couplings are very important issues in
contemporary Higgs physics.

Both of these issues can, at least partially, be addressed by
improving the description of Higgs boson production in gluon
fusion. Indeed, since $gg \to H$ is the main production mechanism of
Higgs bosons at the LHC, a refined understanding of this process in
QCD perturbation theory will lead to an improved understanding of
fiducial cross sections and, eventually, will allow for a better
extraction of various Higgs couplings constants from e.g. Higgs decays
to bosonic final states. 

Although the contributions of bottom and charm loops to the $ggH$ coupling
and direct production of a Higgs boson in quark fusion $q \bar q \to H$, $q \in
\{c,b\}$ are small in the Standard Model, if the Yukawa couplings
differ from their Standard Model values, these light-quark effects in
Higgs production become much more important.  In fact, it was pointed
out \cite{Bishara:2016jga,Soreq:2016rae} that studies of  
kinematic distributions of Higgs bosons produced in hadron collisions   may lead to
interesting constraints on light quark Yukawa couplings, especially at
the high-luminosity LHC.

A particularly important and highly non-trivial kinematic distribution
is the Higgs boson transverse momentum spectrum. At the LHC, Higgs
bosons are produced with very different transverse momenta, from very
small to very large; the $\pt$ distribution peaks at $\pt \approx
15~{\rm GeV}$.  Depending on the value of the Higgs transverse
momentum, the $\pt$ distribution is sensitive to different physics,
from multiple emissions of soft gluons at small $\pt$ to top quark
mass effects at the tail of the spectrum.  The
difficulty in describing the Higgs transverse momentum distribution as a whole is
related to this point.
 
Higgs production in gluon fusion receives contributions from top and
light-quark loops.  Since Yukawa couplings are proportional to quark
masses, top quark loops play the dominant role.  For values of the
transverse momenta $\pt \lesssim m_t$, top loops can be treated in the
$m_t \to \infty$ approximation. This leads to enormous technical
simplifications since, essentially, it allows us to ``remove'' one
loop from the computations that involve the $ggH$ vertex.  As the
result, the $m_t \to \infty$ approximation allowed for the computation
of next-to-next-to-next-to-leading order (N$^3$LO) QCD corrections to
the inclusive cross section and basic kinematic distributions
\cite{Anastasiou:2015ema,Mistlberger:2018etf,Dulat:2017prg}, as well as next-to-next-to-leading 
order (NNLO) QCD corrections to the production of Higgs bosons
in association with one jet
\cite{Boughezal:2015aha,Boughezal:2015dra,Caola:2015wna,Chen:2014gva,
  Chen:2016zka}.  It is also quite straightforward to compute the
${\cal O}(1/m_t)$ corrections to the $m_t \to \infty$ approximation;
they are available for the total cross
section~\cite{Harlander:2009my,Pak:2011hs,Harlander:2009mq} and for the Higgs $\pt$
distribution \cite{Harlander:2012hf,Neumann:2014nha,Neumann:2016dny}. NLO QCD
corrections including the top-quark dependence in the full Standard Model have
become available recently either via a high-energy 
expansion~\cite{Lindert:2018iug,Neumann:2018bsx} or a direct numerical
calculation~\cite{Jones:2018hbb} of the relevant two-loop virtual amplitudes.

When small transverse momenta $\pt \ll m_H$ are considered, radiative
corrections to Higgs production become enhanced by large logarithms
$\ln m_H/\pt$. It is possible to resum these logarithms in case of
Higgs production in gluon fusion if the $ggH$ coupling is point-like,
which is the case in the $m_t \to \infty$ approximation.  Such
resummations were performed with ever increasing accuracy through the
years
\cite{Bozzi:2005wk,deFlorian:2011xf,deFlorian:2012mx,Becher:2012yn}.
The necessary ingredients to compute the
next-to-next-to-next-to-leading logarithmic (N$^3$LL) corrections,
apart from the four-loop cusp anomalous dimension, were obtained in
Refs.~\cite{Catani:2011kr,Gehrmann:2014yya,Li:2016ctv,Vladimirov:2016dll}. This
allowed for a description of the Higgs boson $\pt$ spectrum at
N$^3$LL+NNLO~\cite{Bizon:2017rah}.\footnote{The numerical impact of the
four-loop cusp anomalous dimension is expected to be small~\cite{Becher:2013xia}.}

Resummed computations are usually extrapolated from small transverse
momenta, where they are valid, to large momenta, where they are
matched to fixed-order computations.  As we explained in the previous
paragraphs, the accuracy of both resummed and fixed-order computations
has been constantly increasing; as a result, the $m_t \to \infty$
top-loop mediated contribution to the Higgs transverse momentum
distribution is currently known with a precision of about $10-15$
percent for all values of the Higgs $\pt$
\cite{Boughezal:2015aha,Boughezal:2015dra,Caola:2015wna,Chen:2014gva,Chen:2016zka,
Bizon:2017rah}.

Having reached this level of understanding in the $m_t \to \infty$ limit, 
it is essential to ask if additional  small effects, that
could have been neglected previously, need to be accounted for at the
present level of accuracy and, if so, if they are sufficiently well
understood.  Examples of such contributions are corrections to the $ggH$
interaction vertex due to light quarks and electroweak corrections to the
Higgs transverse momentum distribution.  Both of these effects appear
at the one-loop level; light-quark contributions change the Higgs boson
transverse momentum distribution by about $-5\%$.
For moderate values of transverse momenta, electroweak
contributions to the Higgs $\pt$ spectrum are smaller \cite{Keung:2009bs}
and we neglect them in what follows.

Since QCD corrections in $gg \to H$ are known to be large for the top
quark contribution, it becomes important to understand if a similar
enhancement exists for light quark contributions as well.
Unfortunately, such computations require two-loop calculations with
massive internal particles that are currently hardly feasible.  An
alternative possibility is to compute the corresponding two-loop
amplitudes in the approximation where all kinematic variables and the
mass of the Higgs boson are considered large relative to the quark
mass $m_q$. In this case, one computes a two-loop amplitude as an expansion
in $m_q^2/m_H^2,m_q^2/s,m_q^2/\pt^2$.  In this approximation, the relevant amplitudes 
 have been computed in
Refs.~\cite{Melnikov:2016qoc,Melnikov:2017pgf}.  For $b$-quark loops,
such an expansion is valid for transverse momenta larger than ${\cal
  O}(10-20)$~GeV since corrections to the approximate result for two-loop
amplitudes scale like $(m_b/\pt)^2 \sim 0.2 $ for $\pt = 10~{\rm
  GeV}$. 

Light-quark contributions develop a peculiar double-logarithmic
dependence on the light quark masses $\ln^2(m_H/m_q),
\ln^2(\pt/m_q)$. Such dependences originate from soft quark exchanges
in the loops that facilitate the $ggH$ couplings.  For the processes
$gg \to H+g, qg \to Hq$ and $q \bar q \to H+g$ these terms are
sensitive to gluons emitted from both ``inside'' and ``outside'' the
loops at finite transverse momentum, i.e. to the structure-dependent
radiation.

Light-quark contributions to Higgs production in gluon fusion make the
resummation of the transverse-momentum distribution
difficult~\cite{Grazzini:2013mca,Banfi:2013eda}. Indeed, since both
top and bottom quark loops contribute to the $ggH$ coupling and since
these loops are characterized by very different intrinsic scales for
the structure-dependent radiation ($m_b$ and $m_t$), it appears that
one will have to treat them differently.  However, this is not
possible since the dominant contribution is given by the interference
of the two amplitudes.  In addition, since it is not understood how to
resum the potentially large logarithms $\log(\pt/m_b)$ that appear in
the light-quark loops, it becomes impossible to treat all the
different contributions to the Higgs $\pt$ spectrum on the same
footing. The best thing that one can do is to employ a variety of
prescriptions for combining light quark contributions with small-$\pt$
resummations and to study how the resulting uncertainty in predictions
compares with other sources of theoretical error.

The goal of this paper is to study the Higgs $\pt$ spectrum including
top and bottom-quark contributions at next-to-leading order combined
with next-to-next-to-leading logarithmic transverse momentum
resummation (NLO+NNLL). A similar study at leading order combined with
next-to-leading logarithmic resummation (LO+NLL) was performed in
Ref.~\cite{Grazzini:2013mca}.\footnote{Note that this was referred to
  as NLO+NLL in this reference, while we always use the formal
  accuracy of the differential distribution for the fixed order. In
  the notation of Ref.~\cite{Grazzini:2013mca} our result would be
  NNLO+NNLL.} To this end, we include the recently computed NLO QCD
corrections to light-quark contributions to Higgs production in gluon
fusion \cite{Lindert:2017pky,Melnikov:2017pgf,Melnikov:2016qoc}. We
find that the uncertainty in our matched NLO+NNLL result for the
top-bottom interference contribution to the Higgs transverse momentum
distribution in the region $ 10~{\rm GeV} \lesssim \pt \lesssim 100~{\rm GeV}$
is dominated by ambiguities in the perturbative description of
light-quark loops rather than by uncertainties in the resummation
itself. In particular, we do not find large uncertainties
related to the choice of the resummation scale for the $b$-quark
loops.

The paper is organized as follows. In Section~\ref{sec:resum} we briefly
review the structure of small-$\pt$ resummation for the case of
point-like interactions, and elucidate its main assumptions and
limitations. We also study light-quark
contributions, discuss why in this case the resummation is challenging and
describe a possible pragmatic solution to this problem. In Section~\ref{sec:res}, we explain the
implementation of the resummation procedure for the $b$-quark
contribution and study its ambiguities, and we present
our main results for the Higgs transverse momentum distribution.  We
conclude in Section~\ref{sec:conclusions}.  Some useful formulas and derivations are
collected in the Appendix.

\section{Resummation of the Higgs transverse momentum distribution}
\label{sec:resum}

\subsection{The standard point-like case} 
\label{sec:resgeneric}

We would like to describe the transverse momentum distribution of Higgs 
bosons produced in hadron collisions. This is non
trivial and requires a combination of fixed order and resummed perturbative 
calculations.
Indeed,  depending on the value of the Higgs boson transverse momentum, 
we can distinguish two regions.
For 
large values of transverse momenta $\pt \sim \mh$, one  can compute ${\rm d}\sigma /{\rm d}  \pt$  in 
a perturbative expansion in $\alpha_s$ following standard rules of perturbative Quantum Field Theory.  
For small values of the transverse momentum $\pt \ll \mh$, the situation is different since emerging 
large logarithms  $\ln(\pt/\mh) \gg 1$ may compensate the smallness of the strong coupling 
constant, $\alpha_s \ln^2(\pt/\mh)\sim 1$, and spoil a conventional  perturbative
expansion. To deal with this case, one resums the logarithmically enhanced terms to all orders in the 
coupling constant, and develops a perturbative expansion  {\it on top} of the resummed result. 

Since, eventually, we need to describe the Higgs boson $\pt$ distribution 
for all values of transverse momenta, the two distinct approaches -- resummation and fixed 
order computations -- 
have to be combined.  This is done by smoothly interpolating between results
derived  at small and large $\pt$.  The region where the transition happens is characterized 
by a quantity 
that we refer to as the resummation scale $Q$. This scale has the following physical meaning:
for $\pt \lesssim Q$, the transverse momentum distribution is mostly described by the resummed result,
while for $\pt \gsim Q$ it is mostly described by the fixed order computation.

In order to discuss these concepts more precisely, we consider
the all-order resummation in a toy model, where we work at leading-logarithmic (LL) accuracy. 
To this end, we consider the cumulative distribution 
\be
\Sigma(\pt) = \int \limits_{0}^{\pt} {\rm d} \pt' \; \frac{{\rm d} \sigma}{{\rm d} \pt'}.
\ee
At low $\pt$, we resum the logarithms of $\ln \pt/\mh$ and write 
\be
\Sigma(\pt) =  \Sigma^{\rm resum}(\pt),\;\;\;\; \pt \ll \mh.
\ee
In this region, the distribution is dominated by the emission of
soft and collinear partons. In the LL approximation it is sufficient to
consider the most singular contribution to the QCD matrix elements,
where all final-state partons are soft and strongly ordered in angle.
In this limit, the squared matrix element for the emission of
$n$ extra partons $gg\to H+ n$ is given by the product of the matrix
element for $gg\to H$ times $n$ independent eikonal factors. More specifically, at LL the
partonic $\pt$ distribution 
\begin{align}
\frac{{\rm d} \hat\sigma}{{\rm d}\pt}
  = 
  [d p_H]&\left(\prod_{i=1}^{n} [dk_i]\right)|\mathcal M(p_1+p_2\to H +
  n)|^2 
\notag\\
&\times\delta^{(4)}\left(p_1+p_2 - p_H -\sum_{i=1}^n k_i\right)
\delta\left(\pt-|\sum_{i=1}^{n}\vec{k}_{\perp i}|\right)
\end{align}
can be simplified as
\begin{align}
\label{eq:eikonal}
\frac{{\rm d} \hat\sigma}{{\rm d}\pt}
\simeq [d p_H]\mathcal|\mathcal M(p_1 + p_2\to
H)|^2 \delta^{(4)}(p_1+p_2 - p_H)
\notag\\
\times \frac{1}{n!}\prod_{i=1}^{n}  [dk_i]|M_{\rm soft}(k_i)|^2
\delta\left(\pt-|\sum_{i=1}^{n}\vec{k}_{\perp i}|\right),
\end{align}
where $[dk_i]$, $[dp_H]$ are the phase space volumes of  the $i$-th
parton $k_i$ and the Higgs boson, and $M_{\rm soft}$ is
the matrix element of the single-emission eikonal current. Note that the 
\emph{reduced matrix element $\mathcal M(p_1 + p_2 \to H)$} is
evaluated at \emph{zero transverse momentum}.\footnote{This is valid
  at all logarithmic orders.} 

Starting from Eq.~\eqref{eq:eikonal} it is
possible to show (for details see Appendix~\ref{sec:resummation}) that
the resummed cross section takes the form
\begin{align}
\label{eq:toy}
\Sigma^{\rm resum}(\pt) = \sigma_0 e^{-\int[dk]|M_{\rm
  soft}(k)|^2}\sum_{n=0}^{\infty} \frac{1}{n!}\prod_{i=1}^{n}  \int [dk_i]|M_{\rm soft}(k_i)|^2 \Theta\left(\pt-|\sum_{i=1}^{n}\vec{k}_{\perp i}|\right),
\end{align}
where $\sigma_0$ is the Born cross section for $gg\to H$.
The overall exponential factor contains the all-order effects of
soft-collinear virtual gluons  which are encoded in the
leading divergence of the gluon form factor $\mathcal M(p_1 + p_2 \to
H)$.\footnote{Clearly, all integrals in Eq.~\eqref{eq:toy} are
  divergent in the soft and collinear limits, and require regularization. However, 
 the final result 
  Eq.~\eqref{eq:toy} does not depend on the regularization procedure.}
The distribution
in the 
small $\pt$ region is governed by two competing mechanisms. In the
strict limit $\pt \to 0$, the dominant contribution comes from emissions with finite
transverse momentum $\pt\ll k_{\perp i}\ll \mh$ that mutually cancel in the
transverse plane. This collective effect gives rise to a power
suppressed scaling~\cite{Parisi:1979se}
\begin{equation}
\Sigma(\pt)\sim {\cal O}(\pt^2).
\end{equation}
As $\pt$ increases, but  still remains small compared to $\mh$, the
distribution is described  by kinematic configurations with $\pt\sim k_{\perp i}\ll
\mh$. As discussed in Appendix~\ref{app:LL}, in this region the
cumulative distribution features an exponential suppression of the
form
\begin{equation}
\label{eq:LLexp}
\Sigma(\pt)\sim \sigma_0 \,\exp\left\{-\abar \ln^2 \frac{\mh}{\pt}\right\},
\end{equation}
where $\abar=2C_A \alpha_s/\pi$.

At larger transverse momenta ($\pt\sim \mh$) the approximation that
led to Eq.~\eqref{eq:toy} is not justified anymore. Therefore,
in this region one has to smoothly switch from the resummed prediction
to the fixed-order one, where the effect of the hard radiation is
treated correctly. This can be done for example  using the following
matching formula
\begin{equation}
\label{eq:add_match}
\Sigma(\pt) = \Sigma^{\rm resum}(\pt) + \left ( 
\Sigma^{\rm f.o.}(\pt) - {\cal T}^{\rm f.o.} \left [  \Sigma^{\rm resum}(\pt)  \right ]\right ),
\end{equation}
where we indicate with ${\cal T}^{\rm f.o.}[f]$ the fixed-order
expansion of $f$.  At small $\pt$ the difference between the fixed-order
result and the Taylor expansion of the resummed result is free of
logarithmically-enhanced terms
\begin{equation}
\lim_{\pt \to 0}  \left ( 
\Sigma^{\rm f.o.}(\pt) - {\cal T}^{\rm f.o.} \left [  \Sigma^{\rm resum}(\pt)  \right ]\right )
 = {\rm const}. 
\end{equation}
This allows one to extend the fixed-order description to $\pt\to 0$ and, 
at the same time,  ensures  that  all terms that contain large
logarithms at low $\pt$ are resummed.

The precise way to switch from the resummation to the fixed-order
description is ambiguous. One source of ambiguity comes from choosing
a particular form for the matched cross section (in our example,
Eq.~\eqref{eq:add_match}, we chose to combine the resummed and
fixed-order predictions additively). A second source of ambiguity is
connected with the scale at which the transition from resummed to 
fixed-order result takes place. Although all of these effects are formally of higher-order
both in the resummation and fixed-order counting, their numerical
impact can be non-negligible. We consider the latter issue in what 
follows, while leaving a discussion of the choice of the matching
scheme to the next section.

In order to switch off resummation effects at large $\pt$, 
one can modify the resummed cross section by including controlled
power-suppressed corrections. One possible way to do this is to modify 
the resummed logarithms in Eq.~\eqref{eq:LLexp} as follows\footnote{A
  more correct prescription is to modify the logarithms
  $\ln(\mh/k_{\perp 1})$ where $k_{\perp 1}$ is the transverse momentum of the
  hardest emitted gluon. This technicality is avoided here for the
  sake of clarity, and it will be discussed in
Appendix~\ref{sec:resummation}.}  
\be L\equiv\ln \frac{\mh}{\pt} =
\ln\frac{\mh}{Q}+\ln\frac{Q}{\pt}, 
\ee 
where $Q$ is an arbitrary scale of order $\mh$. Moreover, we write
\be \ln\frac{Q}{\pt}
\to \frac{1}{p}\ln \left(\left(\frac{Q}{\pt}\right)^p+1\right) \equiv \tilde L,
\label{eq:modlog}
\ee
where $p$ is a positive number.
The motivation for the transformations  described above is as follows:
\begin{itemize}
\item First, we split the
resummed logarithm $L$ into the sum of a {\it small} logarithm
$\ln(\mh/Q)$ (with $Q\sim \mh$) and a {\it large} logarithm
$\ln(Q/\pt)$. This operation allows us to introduce a generic scale
$Q$ which then appears in the resummed result. We can now expand $L$
around $\ln(Q/\pt)$, retaining all terms with the desired logarithmic
accuracy. Effectively, this implies that $\ln(\mh/Q)$ is treated
perturbatively at fixed order. In our LL example, for $\pt\sim k_{\perp i}\ll \mh$, this means
\be
\begin{split}
\Sigma^{\rm resum}(\pt) \sim\; 
e^{-\abar L^2} &= 
\exp\left\{-\abar \left[\ln^2 \frac{Q}{\pt} + 2\ln \frac{Q}{\pt} \ln \frac{\mh}{Q}
 + \ln^2 \frac{\mh}{Q}\right]\right\} \\
&= \exp\left\{-\abar \ln^2 \frac{Q}{\pt} + \mathcal O\left( \alpha_s L
  \right)\right\}\left(1-\abar \ln^2 \frac{\mh}{Q}+{\cal
    O}(\alpha_s^2)\right)\\
&\simeq \exp\left\{-\abar \ln^2 \frac{Q}{\pt}\right\},
\end{split}
\label{eq:mod_log_ll}
\ee
where all terms beyond LL were neglected. This prescription is
convenient because the $Q$-dependence is always of
higher-logarithmic order and, therefore,  a $Q$-variation probes the
size of subleading logarithms that are not considered in the resummation.
\item Second, we modify the logarithm 
$\ln(Q/\pt)$ by including power-suppressed terms that force 
  $\tilde{L}$ to vanish at large $\pt$. These modifications do not 
  affect the small-$\pt$ limit. Indeed, it follows from Eq.~\eqref{eq:modlog} that
\begin{align}
\label{eq:scaling}
\tilde{L} \sim \ln\frac{Q}{\pt},\,\,{\rm for}\,\,\pt\ll Q;\qquad \tilde{L} \sim \frac{1}{p}\left(\frac{Q}{\pt}\right)^p,\,\,{\rm for}\,\,\pt\gg Q.
\end{align}
As a consequence, the {\it resummation} scale $Q$ and the scaling
parameter $p$ must be chosen in such a way that the high-$\pt$ scaling of the
resummed component (and its fixed-order expansion) does not modify the
scaling of the fixed-order prediction. This means that  $p$ and $Q$ are 
 to be chosen
in such a way that the resummed component vanishes more quickly than
the fixed-order result for $\pt\gsim Q$.
\end{itemize}

\begin{figure*}[htb]
\centering
\hspace{-0.9cm} \includegraphics[width=.5\textwidth]{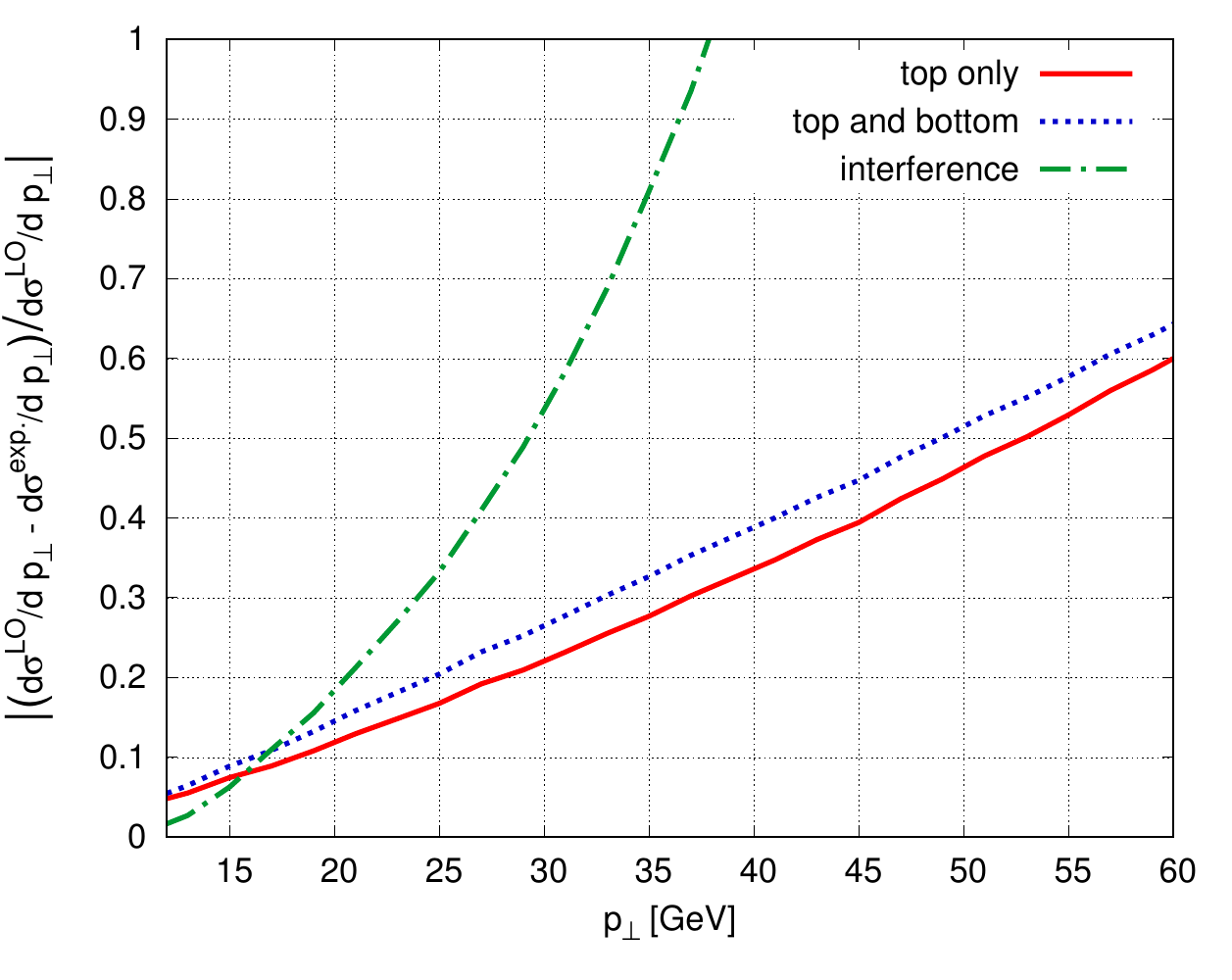} 
\caption{Differences between leading-order distributions and their 
  logarithmic contributions, normalized to the leading-order results. 
 The three curves show the case for top quark (solid/red), top
  and bottom quarks (dotted/blue), and top-bottom interference 
  (dot-dashed/green). See text for details.}
\label{fig:scale_choice}
\end{figure*}

The above discussion shows that $Q$ is indeed the scale at
which the transition between resummed and fixed-order results 
occurs. Similarly to the renormalization and factorization
scales, its choice is ambiguous, although certain conditions should be satisfied. 
Indeed, it is clear that (a) $Q$ should not
be too different from $\mh$, to ensure that $\ln Q/\mh$ are not large
and (b) it should approximately correspond to the scale at which the soft
and collinear approximations to the matrix element and kinematics
break down. In practice, one can choose $Q$ by comparing the exact
result $\Sigma^{\rm f.o.}$ with the expansion of the resummed result
$\mathcal T^{\rm f.o.}[\Sigma^{\rm resum}]$, and set $Q$ to the $\pt$
scale at which the two start to significantly deviate  from each other. This is
illustrated in Figure~\ref{fig:scale_choice}, which shows the
difference between the LO differential $\pt$
spectrum and the expansion of the resummed result at the same order. Specifically,
we plot 
\begin{equation}
\bigg|\left(\frac{d \sigma^{\rm LO}}{d\pt}-{\mathcal T}^{\rm LO}
\left[\frac{d\sigma^{\rm resum}}{d\pt}\right]
\right)\bigg/\frac{d \sigma^{\rm LO}}{d\pt}\bigg|.
\end{equation}
We observe that when only the top contribution is included (solid, red curve),
the logarithmic terms account for about half of the fixed-order result at
scales $\pt\sim 50 - 60$\,GeV. This suggests that the resummation
scale should be of this order. We conventionally choose $Q=\mh/2$ as a
central value. As far as the choice of the parameter $p$ is concerned,
we have to ensure that at large $\pt$ the resummed component vanishes
faster than the fixed order. Considering the asymptotic scaling in
Eq.~\eqref{eq:scaling}, we choose $p=4$ which guarantees that the
differential distribution vanishes as fast as $1/\pt^5$ for
$\pt\gsim Q$. In principle, any value of $p$ greater than $3$ will
equally do, since $p$ only determines how fast the resummation is
turned off above the scale $Q$. We have indeed checked that by varying
$p$ by one unit around $p=4$ the results do not change significantly.

The same figure shows results for the full spectrum where 
both top and bottom loops are included (dotted, blue curve), and results for 
top-bottom interference
(dot-dashed, green curve).\footnote{We ignore the bottom squared contribution, which is
  completely negligible in the Standard Model.} In this case the situation changes
considerably, and this will be the subject of the next section.

\subsection{Issues with $b$ quarks}
\label{sec:res_bquark}
The ``standard'' approach to resummation described in
Section~\ref{sec:resgeneric} becomes problematic in case of the Higgs
boson production in hadron
collisions~\cite{Mantler:2012bj,Grazzini:2013mca,Banfi:2013eda,Hamilton:2015nsa,Bagnaschi:2015qta,Bagnaschi:2015bop}.
The difficulty is related to the fact that the $ggH$ vertex {\it is
  not point-like} but, rather, is induced by a quark loop.  The
presence of the quark loop implies the existence of
structure-dependent radiation with peculiar properties and has
important consequences for the resummation.  The key to the following
discussion is the appreciation of the fact that the
structure-dependent radiation is suppressed if $\pt$ is smaller than
the mass of the quark but it becomes important otherwise. For $\pt$
larger than the quark mass, the soft and collinear approximations that
provide the foundation for small $\pt$ resummation become unreliable,
as they focus on emissions off external lines and systematically
neglect structure-dependent effects. In this section we elaborate on
this issue.

We consider Higgs boson production in gluon fusion mediated by a quark loop.
We denote the mass of the quark by $m_q$ and consider two cases
$m_q \gg \mh$ and $m_q \ll \mh$.  In the first case, the structure
dependence enters at $\pt \gsim m_q \gg \mh$, so that emissions off
the external lines dominate for transverse momenta up to the Higgs
mass and even higher. Therefore, if we restrict ourselves to values of
$\pt$ that are comparable to $\mh$, the situation is no different from
point-like interaction, and there are no issues in the resummation
procedure described in the previous section. In the Standard Model,
this is indeed what happens with the top loop contribution to the Higgs boson
transverse momentum spectrum.

The second case, $m_q \ll \mh$ is very different.
Indeed, in this case there are three distinct regions 
$\pt \lsim m_q$, $m_q \lsim \pt \lsim \mh$ and
$\mh \lsim \pt$.  In the first region $\pt \lsim m_q$, the transverse
momenta of the Higgs boson and the recoiling partons are typically
small enough not to resolve the structure of the loop
and the extra radiation factorizes. For the bottom-quark
contribution ($m_q\sim 5$\,GeV) the effect of additional  QCD
radiation is strongly  suppressed in this region by all-order effects, so that its impact
on  the total cross section is small.\footnote{This
  region is also very sensitive to non-perturbative effects.} Note, however,  that in
this region there are large logarithmic contributions of the type
$\ln^2 \mh/m_b $, whose resummation is not fully understood
even at the lowest perturbative order~\cite{Melnikov:2016emg,Liu:2017vkm}. 

In the second region $m_q \lsim \pt \lsim \mh$ the structure-dependent
radiation becomes essential and the $ggH$ vertex does not
factorize. In addition to the usual logarithms $\ln \mh/\pt$, the
radiation gives  rise to logarithms $ \ln \pt/m_q$ and
$\ln \mh/m_q$, whose origin and potential resummation are not well
understood.\footnote{For some recent studies, see
  e.g. Refs.~\cite{Melnikov:2016emg,Liu:2017vkm, Caola:2016upw}.} 

The reason why 
the small-$\pt$ resummation is problematic in this region is the following.
Emissions off internal lines can
become as important as emissions off external lines, and together they
probe the loop structure of the $ggH$
vertex. It follows that approximating the small $\pt$ region with an
on-shell $ggH$ form factor is not justified.
In particular, while form factor effects in the top-quark case
only introduce  $(\pt/\mh)$-suppressed corrections, in this case they
both introduce a new logarithmic structure ($ \ln \pt/m_q$ and
$\ln \mh/m_q$) and suppress radiation with 
$\pt\gsim m_q$. In other words, while in the top case, described in
Section~\ref{sec:resgeneric}, at finite $\pt$, the
coefficients of the logarithms differ from the resummed result by
$\pt$-suppressed terms, in the $b$-quark case, this
difference contains new logarithmic terms $ \ln \pt/m_b$ and
$\ln \mh/m_b$ in the region $m_b \lsim \pt \lsim \mh$.

As a consequence, the collinear
approximation should not be expected to work far away from the
$b$-quark threshold. To quantify this effect, we go back to
Figure~\ref{fig:scale_choice}. We see that, while for the top-only case
(solid, red line) the collinear approximation to the leading order
accounts for half of the result at about $\pt\sim 50-60$\,GeV, for the
top-bottom interference this scale is  reduced to about
$30$\,GeV. When top and bottom contributions are considered
together, this effect is less dramatic since  in the SM the
interference accounts for about  $\sim 5\%$ of the full
result. This can be seen in the dotted blue curve of
Figure~\ref{fig:scale_choice}.

Because of the above issues, it is clear that constructing a
reasonable description of the $b$-quark contribution to the Higgs
transverse momentum distribution is problematic.  Since, as we already stressed, in this case
the resummation of potentially large logarithms is not entirely
understood, the best we can do is to use different ways to
interpolate between regions of small and large transverse momenta and check to what extent 
the different results
are compatible.

As already stated, the Higgs boson production in the Standard Model is dominated
by the top quark loop; the bottom loop provides a very small
contribution that is lifted up to ${\cal O}(-5\%)$ by its interference
with the top amplitude.  Because of this, a ${\cal O}(20-30\%)$ control on
the top-bottom interference is sufficient to control the Higgs transverse
momentum spectrum at the few percent level.  With this in mind, we now
study in more detail the  different ways to treat the bottom contribution.

One option is to apply Eq.~\eqref{eq:add_match} with the resummation
scale set to $Q \sim m_b$~\cite{Grazzini:2013mca}. This choice is equivalent 
to employing  fixed-order description for all values of transverse momenta. 
Indeed,  for $1~{\rm GeV}  \lsim p_\perp \lsim Q \sim m_b$, the 
resummed logarithms $\ln(Q/\pt)$ never become large and 
for $\pt \gsim Q$ the fixed-order
result is adopted anyhow.  Since a typical error made within this
approach is provided by uncalculated higher order terms, if we use
a NLO computation for the
interference, we make an error of order\footnote{Note that these
  estimates refer to the top-bottom \emph{interference} contribution.
  As we said, the term proportional to $y_b^2$ is negligible in
  the Standard Model.}  $[\alpha_s/(2\pi)]^2 \ln^4(\mh/m_b)$ and
$[\alpha_s/(2\pi)]^2 \ln^4(\pt/\mh)$ which both evaluate to $15-20$
percent, for $\pt \sim m_b \sim 5~{\rm GeV}$.

The previous option amounts to neglecting the resummation for
the top-bottom interference and to using the fixed-order result for all
transverse momenta; the other extreme alternative consists of
extending the resummation beyond its established domain of validity.  We
can do this by using the same resummation scale,  $Q \sim \mh/2$,  
both for the top and the top-bottom 
interference contributions~\cite{Mantler:2012bj,Banfi:2013eda}. 
In this case, at
low $\pt$ we introduce logarithms $\ln \pt/Q$ in the interference
through the resummation prescription which are not guaranteed to be
correct and, by doing that, we effectively introduce errors that are
similar to those discussed above. At higher $\pt$ the impact of
these logarithms becomes smaller since at $\pt \sim Q$ the resummation
effects smoothly turn off and we recover the fixed order prediction.
Hence, when we under-
or over-resum logarithms 
we expect comparable ${\cal O}(20\%)$  theoretical errors on the interference contribution
to the Higgs $\pt$ spectrum. 
An
important question is whether these different sources of uncertainties pull
the predictions apart or they remain compatible with each other.

Before concluding this section, we mention that within the
\emph{additive} matching scheme of Eq.~\eqref{eq:add_match} the
resummation term, which is proportional to the lowest-order form factor at zero transverse
momentum,  is added to  the fixed order result.
As we mentioned earlier, the form factor effects lead to 
a dependence  of the leading order amplitude on  $\pt$, that  is not captured
in this approach. To account for this, we also
consider a \emph{multiplicative} matching scheme, which can be schematically
defined as\footnote{The actual implementation of this procedure requires
  extra care, as described in Appendix~\ref{sec:matching}.}  
\be \Sigma(\pt) =
\Sigma^{\rm resum}(\pt) \,\mathcal T^{\rm f.o.}\left[\frac{\Sigma^{\rm f.o.}(\pt)}{\Sigma^{\rm
    resum}(\pt)}\right].
\label{eq:mult_match}
\ee
Similarly to  Eq.~\eqref{eq:add_match}, Eq.~\eqref{eq:mult_match} smoothly interpolates between  a 
low $\pt\ll Q$ region, where resummation dominates,  to a large $\pt\gg Q$ region, where the
result is obtained from a fixed order calculation. Clearly,  the fixed
order accuracy is preserved in the $\pt\to 0$ limit. The main difference with 
Eq.~\eqref{eq:add_match} is that now the higher order terms induced by the resummation
in the transition region are weighted with the fixed (lower) order result at \emph{finite}
transverse momentum. This should at least partially capture the $\pt$ dependence of the
exact higher order amplitude, and lead to a more realistic description of the physics. 
Because of this, we choose the multiplicative matching scheme 
Eq.~\eqref{eq:mult_match}  as our default matching
scheme.
 Nevertheless, matching ambiguities are by construction of higher-order
nature and, therefore, any matching prescription is formally equally
valid. Differences between matching prescriptions can be used to
estimate the uncertainty in the transition region.

\section{Results}
\label{sec:res}

\subsection{Inclusion of bottom-quark loops and matching uncertainties}
\label{sec:method_bquark}

In this section we describe the practical and technical implementation of the
top-bottom interference in the resummation and matching, and the
uncertainty associated with it.  As we described in 
Section~\ref{sec:res_bquark}, the rigorous resummation in the presence of the bottom-quark
loop is currently impossible. To remedy this problem, we adopt
different approaches to include this contribution in the matched
result. We use the arbitrariness in the 
choice of  the resummation scale associated with the
top-bottom interference and in the choice of the  matching scheme
to assess the inherent ambiguity of the resummation procedures.

We start by discussing  the resummation scale. We treat
separately the contribution of the top-squared amplitude and the
top-bottom interference.\footnote{For our numerical results, we also
  include the bottom squared contribution, which is however negligible
  in the Standard Model.} In particular, we associate two different resummation
scales with the top and the interference contributions, and we use
the following notation to denote the various cumulative distributions
\begin{align}
\Sigma_{t+b}(\pt,Q_t,Q_b) &\to {\rm top~and~bottom,~including~the~interference};\\
\Sigma_{t}(\pt,Q_t) &\to {\rm only~top}.
\end{align}
As explained in Section~\ref{sec:resgeneric}, for the top-only
contribution we set the resummation scale to $Q_t=\mh/2$. For
the interference, instead, we use the following  prescriptions
to quantify the associated uncertainty (see Section~\ref{sec:res_bquark}):
\begin{itemize}
\item We choose $Q_b \sim m_b$, effectively switching  off the resummation 
for the interference at scales of the order of the bottom mass.
As it was initially suggested in Ref.~\cite{Grazzini:2013mca}, we choose $Q_b=2 m_b$ as our central scale. This is achieved by computing
\begin{equation}
\Sigma_{t+b}(\pt,\mh/2,2\mb) = \Sigma_{t}(\pt,\mh/2) + \Sigma_{t+b}(\pt,2\mb,2\mb) - \Sigma_{t}(\pt,2\mb).
\end{equation}
This implies that in the region of transverse momenta that we are interested in, the
interference is described only at fixed order and no
resummation for this contribution is performed.

\item We consider the opposite situation in which we rely on the
  collinear approximation also for $\mb \ll \pt$, and simply treat the
  new logarithmic terms that appear above this scale as a regular
  remainder that can be  described at fixed order. As a consequence,
  the resummation for the interference contribution is switched off,
  as in the top-only case, at scales of order $60$\,GeV. We 
   choose $Q_t=Q_b=\mh/2$ as our central scale, for simplicity.
\end{itemize}

In both approaches, logarithms of the ratio $\pt/\mb$ are not
resummed. Although in the region $\mb \ll \pt \ll \mh$ these
logarithmic terms can be potentially large and therefore should be
included to  all orders, recent studies seem to suggest that an
accurate prediction of these terms is achieved by considering the
first few terms in the fixed-order perturbative
expansion~\cite{Melnikov:2016emg,Liu:2017vkm}.

As far as the resummation is concerned, the result will be
nearly identical to the $m_q\to\infty$  one.\footnote{\emph{\rm HEFT} in the
  following.} The only difference is that now the Born squared
amplitude and the hard-virtual correction will contain the
full dependence on the top and bottom masses. In particular, no
modification of the $\pt$-dependent radiation pattern is
introduced. Technically, we implement the LO and the NLO amplitudes for $gg\to H$ with
full mass dependence following Ref.~\cite{Aglietti:2006tp}.

We now study numerically the difference between the two prescriptions for the
bottom resummation scale. We start by introducing the setup that we
adopt for our predictions. We consider proton collisions at the
$13~{\rm TeV}$ LHC.  The Higgs boson mass is taken to be $m_H =
125~{\rm GeV}$ and the top and bottom pole masses\footnote{We work in the
on-shell renormalization scheme as a default.} are set to $m_t=173.2~{\rm GeV}$ and $m_b=
4.75~{\rm GeV}$, respectively.  We work within a fixed flavor-number
scheme ($n_F=5$) and use the PDF4LHC15\_nnlo set of parton
distribution functions \cite{Butterworth:2015oua} interfaced through
LHAPDF6~\cite{Buckley:2014ana}. 
We use the value of the strong coupling constant $\alpha_s$ provided by the
PDF set.  As central values for the renormalization and factorization
scales we take
\begin{equation}
\mu_R = \mu_F = M_T/2, \;\;\; \text{with} \;\;\; M_T = \sqrt{m_H^2 + p_{\perp}^2} .
\end{equation}

In order to estimate the perturbative uncertainties for each
prediction, we perform a $7$-point variation of the factorization
($\mu_F$) and renormalization ($\mu_R$) scales around the central
value by a factor of two. Moreover, we vary $Q_t$ and $Q_b$ by a
factor of two around their respective central values, keeping fixed
$\mu_R=\mu_F=M_T/2$. The final uncertainty band is obtained as the
envelope of all above variations. As a default, we
adopt the multiplicative scheme discussed in
Section~\ref{sec:res_bquark} and described in detail in
Appendix~\ref{sec:matching}.

The fixed-order NLO results for the top-bottom interference are based
on the calculation presented in \cite{Lindert:2017pky}, which in turn
comprises the two-loop amplitudes for $gg\to H g$, $q\bar q \to H g$
and $qg \to H q$ derived in
\cite{Melnikov:2016qoc,Melnikov:2017pgf} together with corresponding
loop-squared real radiation amplitudes as provided by {\sc
  OpenLoops}~\cite{Cascioli:2011va,OLhepforge} combined with {\sc
  Collier}~\cite{Denner:2016kdg}. For the Monte Carlo integration and
subtraction the {\sc Powheg-Box-Res} is
used~\cite{Alioli:2010xd,Jezo:2015aia}.

\begin{figure*}[htb]
\centering
\hspace{-0.1cm}\includegraphics[width=.5\textwidth]{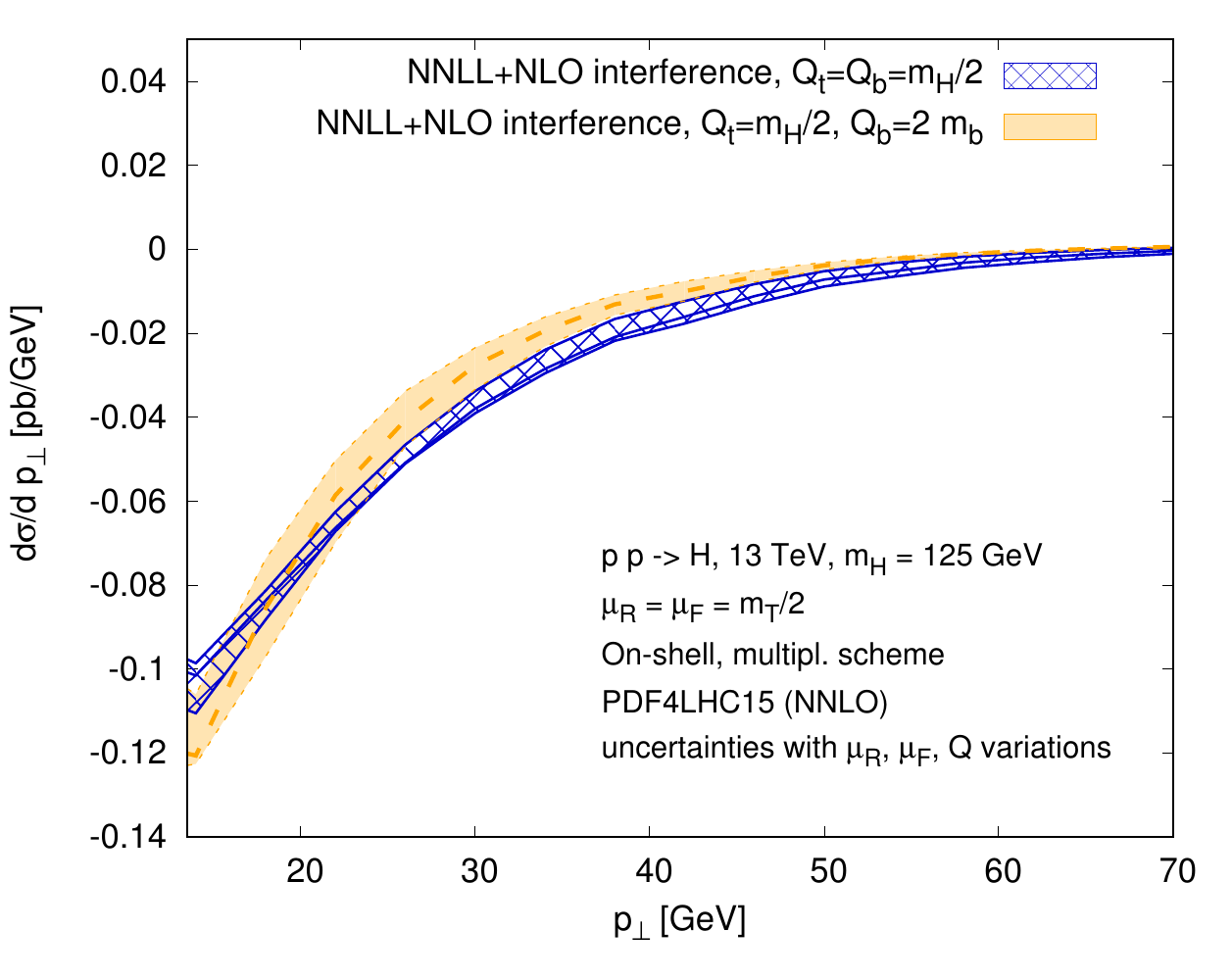} 
\hspace{-0.1cm}\includegraphics[width=.5\textwidth]{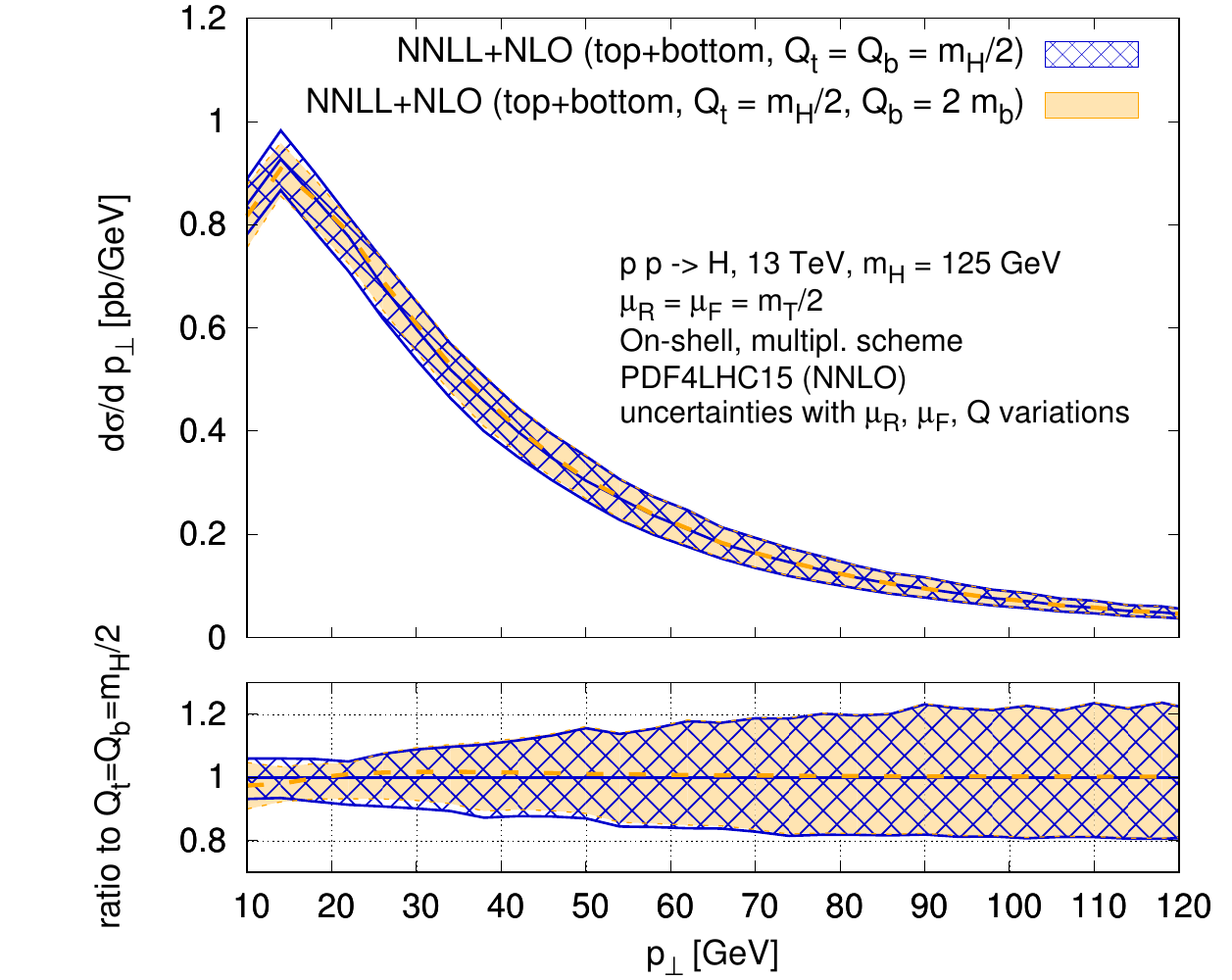} 
\caption{Comparison between two resummation-scale prescriptions for top-bottom interference (left) and full (right) distributions. See text for details.}
\label{fig:two-scales-comparison}
\end{figure*}

We now discuss the dependence on the choice of the resummation scale associated with the bottom contribution. We start by comparing results for  the top-bottom interference 
for two values of the resummation scale $Q_b$. The results are displayed 
in the left plot in Figure~\ref{fig:two-scales-comparison}.
The two predictions differ by about $20\%$ for $\pt\sim30$\,GeV, in
line with what we expected from the discussion in
Section~\ref{sec:res_bquark}. We note, however, that although the two
results are computed for very different choices of the resummation
scales, they are still compatible (although marginally for
$\pt\gsim25$\,GeV) within their respective uncertainties.  Only for
$\pt\gsim 50$\,GeV the two results differ significantly, since the
interference obtained with $Q_b=2\,\mb$ vanishes faster than the one
obtained with $Q_b=\mh/2$. However, in this region the contribution of
the interference to the physical spectrum is completely
negligible. Each of the two results has a relative uncertainty of
about $15\%$ for $\pt\lsim 40$\,GeV. The variations of the resummation
scales around their central value, and the variation of $\mu_R$ and
$\mu_F$ have a similar impact on the final band.

The right plot of Figure~\ref{fig:two-scales-comparison} shows an
analogous comparison for the transverse momentum distribution that  
includes both top and bottom contributions.  Since the interference
only accounts for about $5\%$ of the full result, we find that the two
resummation prescriptions for the top-bottom interference 
are  indistinguishable within the
uncertainties of the top contribution. Indeed, in this case the uncertainty band is dominated by
the $\mu_R$ and $\mu_F$ variation of the top contribution, which
amounts to about $10-15\%$ for $\pt\lsim 40$\,GeV, while the
resummation-scale uncertainty amounts to about $5\%$ in this region.
Note however that the top-only contribution has been computed one
order higher, both in 
fixed-order QCD~\cite{Chen:2014gva,Chen:2016zka,Boughezal:2015aha,Boughezal:2015dra,Caola:2015wna}
and in the resummation framework~\cite{Bizon:2017rah}.  In this paper, we
focus on the $b$-quark effects and hence do not  include these results but, as a matter of principle, 
they can be used to further reduce the uncertainty on the top contribution. 

We now investigate the second source of resummation ambiguity, namely
the choice of the matching scheme. As discussed in
Section~\ref{sec:res_bquark}, besides our default multiplicative
scheme we also consider an additive scheme. Both schemes are
precisely defined in Appendix~\ref{sec:matching}. We remind the reader
that, as far as the top-bottom interference is concerned, the main
qualitative difference between the two approaches is that within the
\emph{additive} matching scheme, the resummation contribution 
is proportional  to the $gg \to H$ form factor at zero transverse momentum whereas 
in the \emph{multiplicative} matching scheme it is weighted by the form factor 
$g^* g^* \to H$ at finite transverse momentum. 
In order to study this source of ambiguity more precisely, we consider
the additive matching scheme, with two different scales ($Q_b=\mh/2$
and $Q_b=2\mb$), and compare the results to the multiplicative scheme.

\begin{figure*}[tb]
\centering
\includegraphics[width=.48\textwidth]{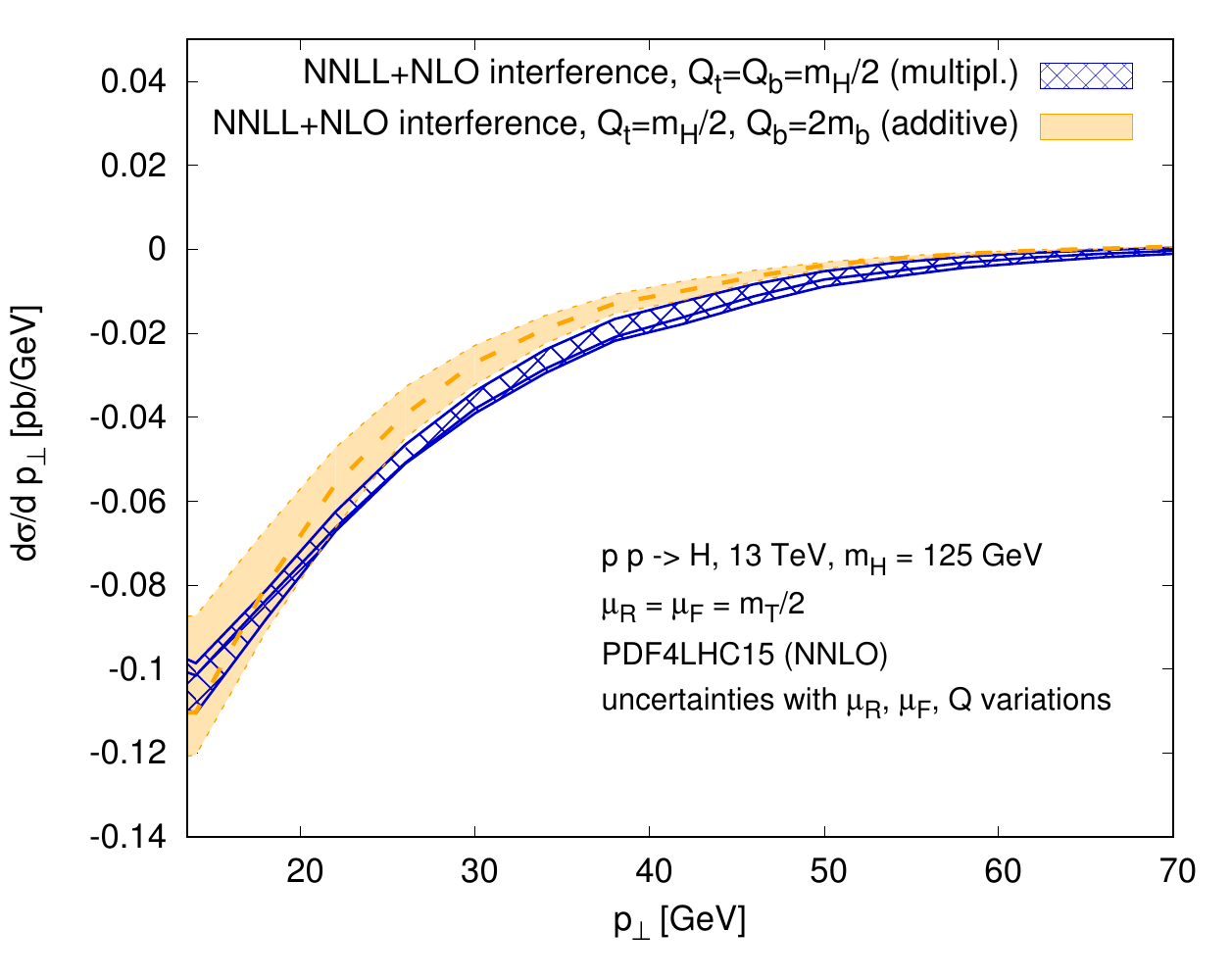}
~
\includegraphics[width=.48\textwidth]{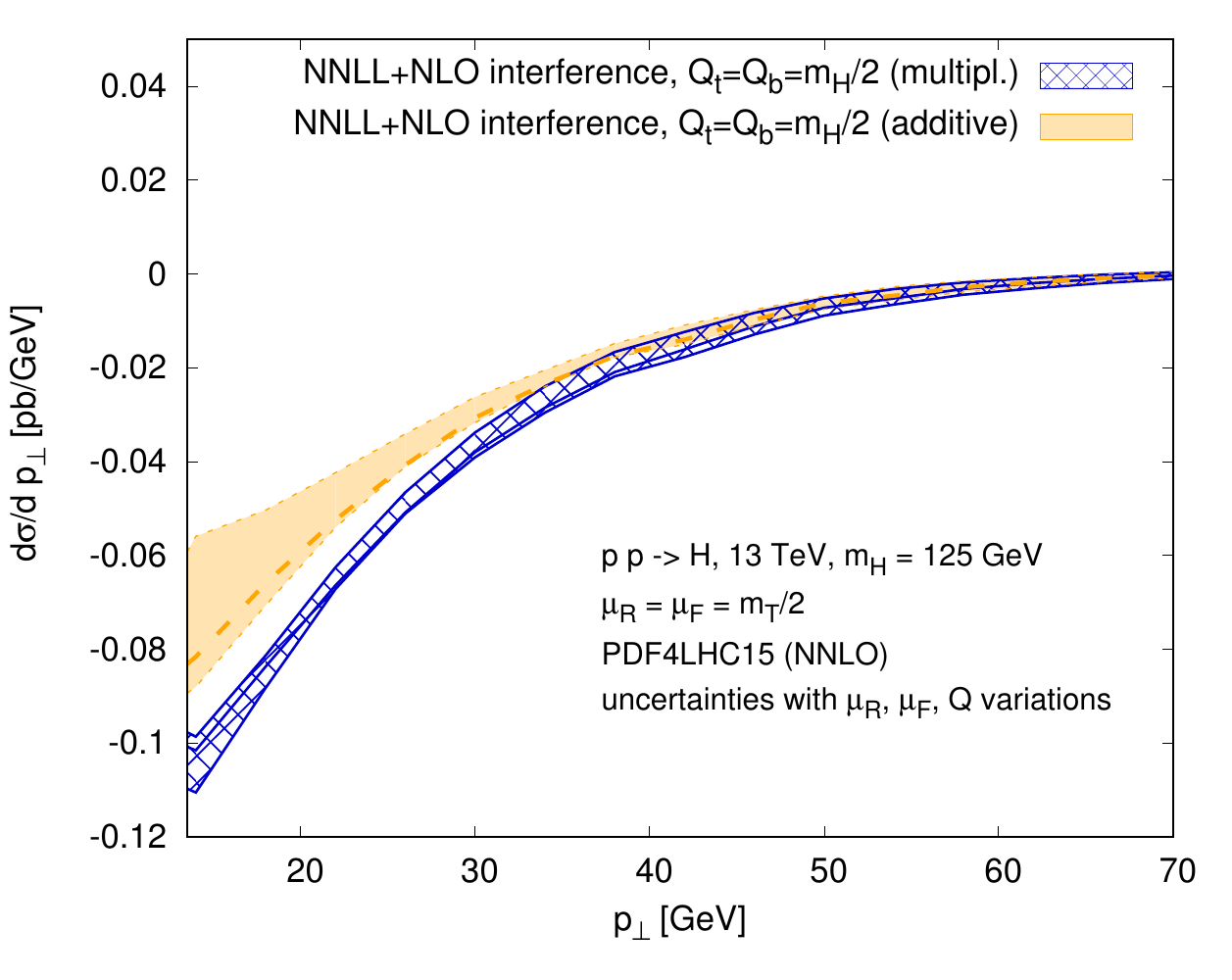}
\caption{Comparison between the additive scheme with $Q_t=\mh/2$,
  $Q_b=2\mb$ (left plot) or $Q_t=Q_b=\mh/2$ (right plot), and the
  default multiplicative scheme with $Q_t=Q_b=\mh/2$. }
\label{fig:schemes-comparison-interference}
\end{figure*}

\begin{figure*}[tb]
\centering
\hspace{-0.9cm} \includegraphics[width=.5\textwidth]{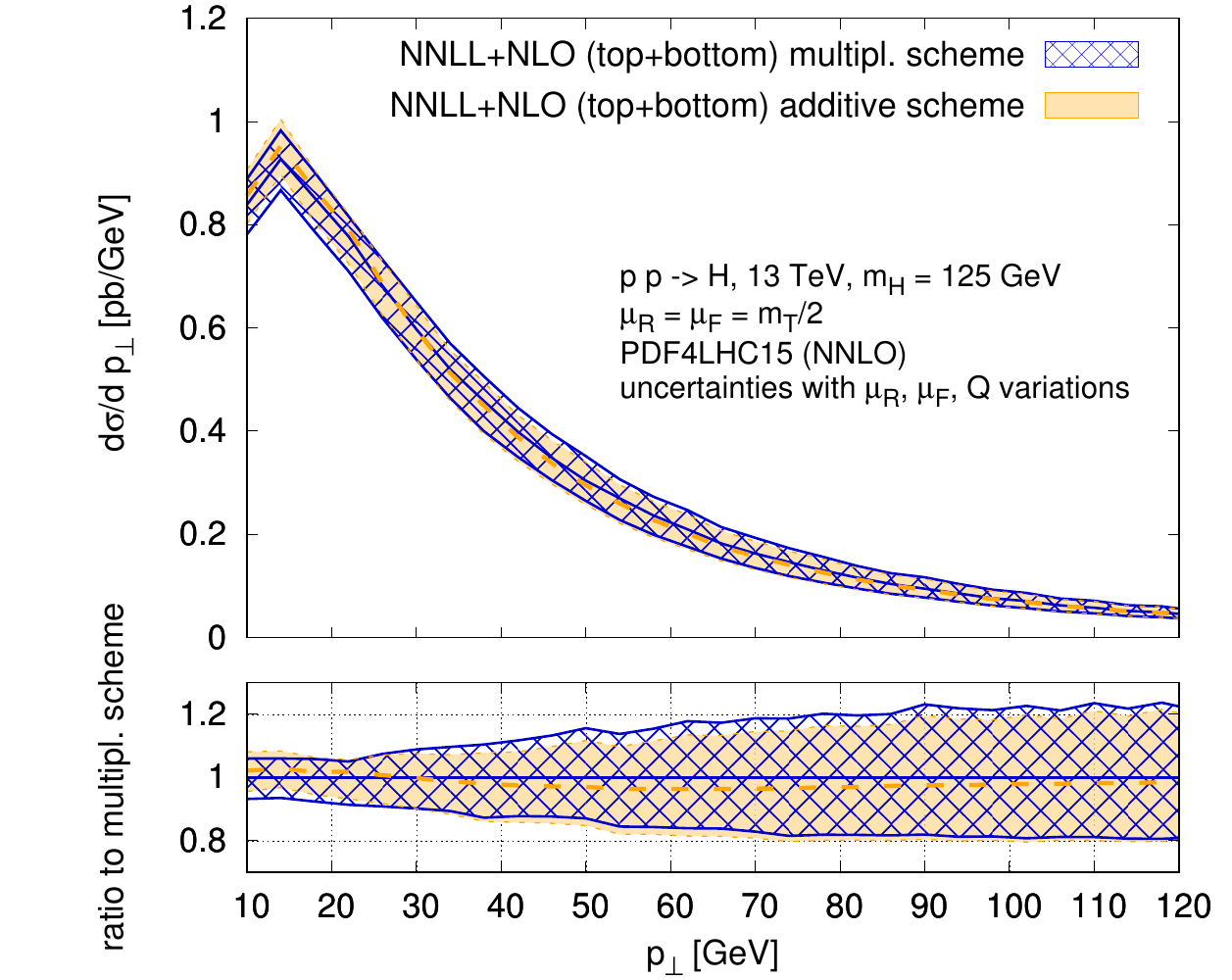} 
\includegraphics[width=.5\textwidth]{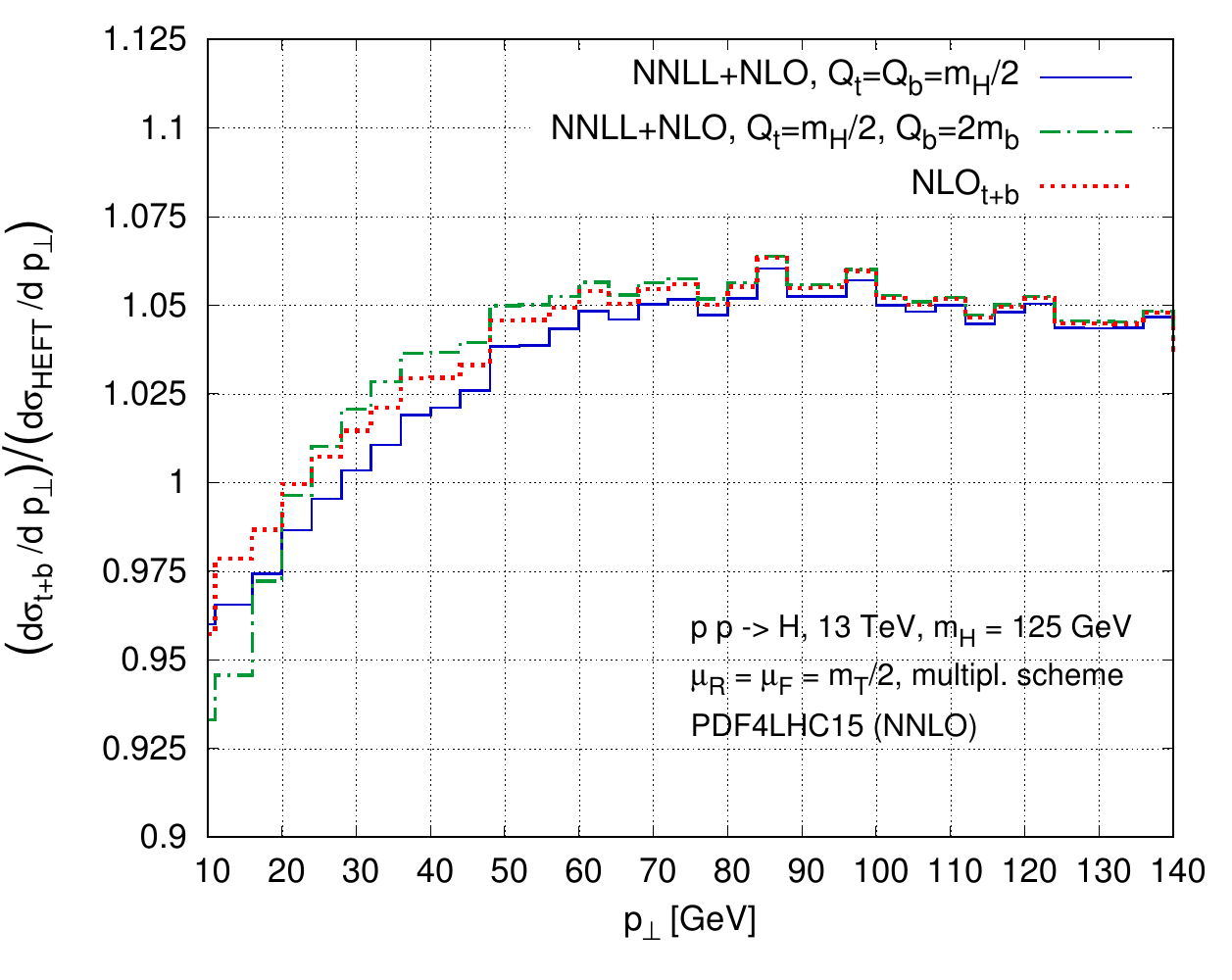}
\caption{Left plot: comparison for the full distribution between the additive scheme with $Q_t=Q_b=\mh/2$, and the default multiplicative scheme with $Q_t=Q_b=\mh/2$. Right plot: ratio of the full distribution with different accuracies to the HEFT prediction based on the same accuracy. In the right plot all matched results are obtained within the multiplicative matching scheme.}
\label{fig:schemes-comparison-full}
\end{figure*}

Since in the additive scheme the resummed contribution does not
include form-factor effects,  we expect sizable
differences between results obtained with $Q_b=\mh/2$ and
$Q_b=2\mb$. We recall that this is not the case in the multiplicative
scheme (see Figure~\ref{fig:two-scales-comparison}) where form-factors
effects are automatically accounted for. We then show, in
Figure~\ref{fig:schemes-comparison-interference}, the comparison
between the top-bottom interference in the default multiplicative
scheme with $Q_b=\mh/2$ and the additive scheme with $Q_b=2\mb$
(left plot) and $Q_b=\mh/2$ (right plot).
We observe that the difference between the two schemes is larger when
the additive scheme with $Q_b=\mh/2$ is used. Nevertheless, we find 
that also in this case, the difference between the
two schemes for the interference does not
exceed $\sim 20\%$ in the bulk of the distribution. Again, 
the full transverse momentum distribution, shown in the left plot
of Figure~\ref{fig:schemes-comparison-full}, is only mildly affected
by this ambiguity.  

Finally, in the right plot of
Figure~\ref{fig:schemes-comparison-full}, we show the ratio of the
full distribution computed using the default multiplicative scheme, to the
corresponding HEFT result. The default result, i.e. multiplicative
matching scheme with $Q_t=Q_b=\mh/2$, is in good agreement
with the NLO prediction. For comparison, we also report the other
extreme solution obtained with the multiplicative scheme with
$Q_t=\mh/2$, $Q_b=2\mb$. We observe that
this choice is in good agreement with both the fixed order and the
default matched solution.

In summary, we find that a conservative approach towards the
inclusion of bottom-mass effects in the matched prediction for the
Higgs-$\pt$ spectrum leads to a $\sim 20\%$ ambiguity on the
top-bottom interference in the region $\mb\lsim \pt$. Since the interference 
provides a rather small contribution to the Higgs transverse momentum 
distribution, this ambiguity translates
into a few-percent uncertainty in the Higgs  $p_\perp$ spectrum.

In what follows, we will use the result obtained with the multiplicative
matching scheme with $Q_t=Q_b=\mh/2$ as our central value. To estimate
uncertainties, we will consider the envelope of scale variations and
the result obtained either by using the multiplicative or the
additive scheme with $Q_t=\mh/2$, $Q_b=2\mb$. 
In addition to these source of uncertainty, an
additional ambiguity arises from the choice of the renormalization
scheme for the quark masses. This will be discussed in the next
section.

\subsection{Mass-scheme uncertainty and final results}
\label{sec:results}

In this section, we present our final results for the NNLL+NLO
matched distributions. We use as default the
multiplicative matching scheme with resummation scales $Q_t=Q_b=\mh/2$.
We renormalize the bottom-quark mass in the on-shell scheme.
To estimate the uncertainty we change the details of the resummation and matching
as explained in the previous section. In addition, we consider 
a different renormalization scheme for the bottom quark mass to estimate 
the related uncertainty.  To this end, we employ the $\overline{\rm MS}$ renormalization scheme. 
We take the mass renormalization scale to be $100~{\rm GeV}$, and use
$m_b = m_b^{\overline {\rm MS}}(100~{\rm GeV})~=~3.07~{\rm
  GeV}$ as an input parameter.\footnote{We calculated this value using
  the program RunDec \cite{Chetyrkin:2000yt,Herren:2017osy} with the
  input value $ m_b^{\overline{ \rm MS}} ( m_b^{\overline{ \rm MS}}
  )~=~4.2~{\rm GeV}$.} 
\begin{figure*}[tb]
\centering
\includegraphics[width=.48\textwidth]{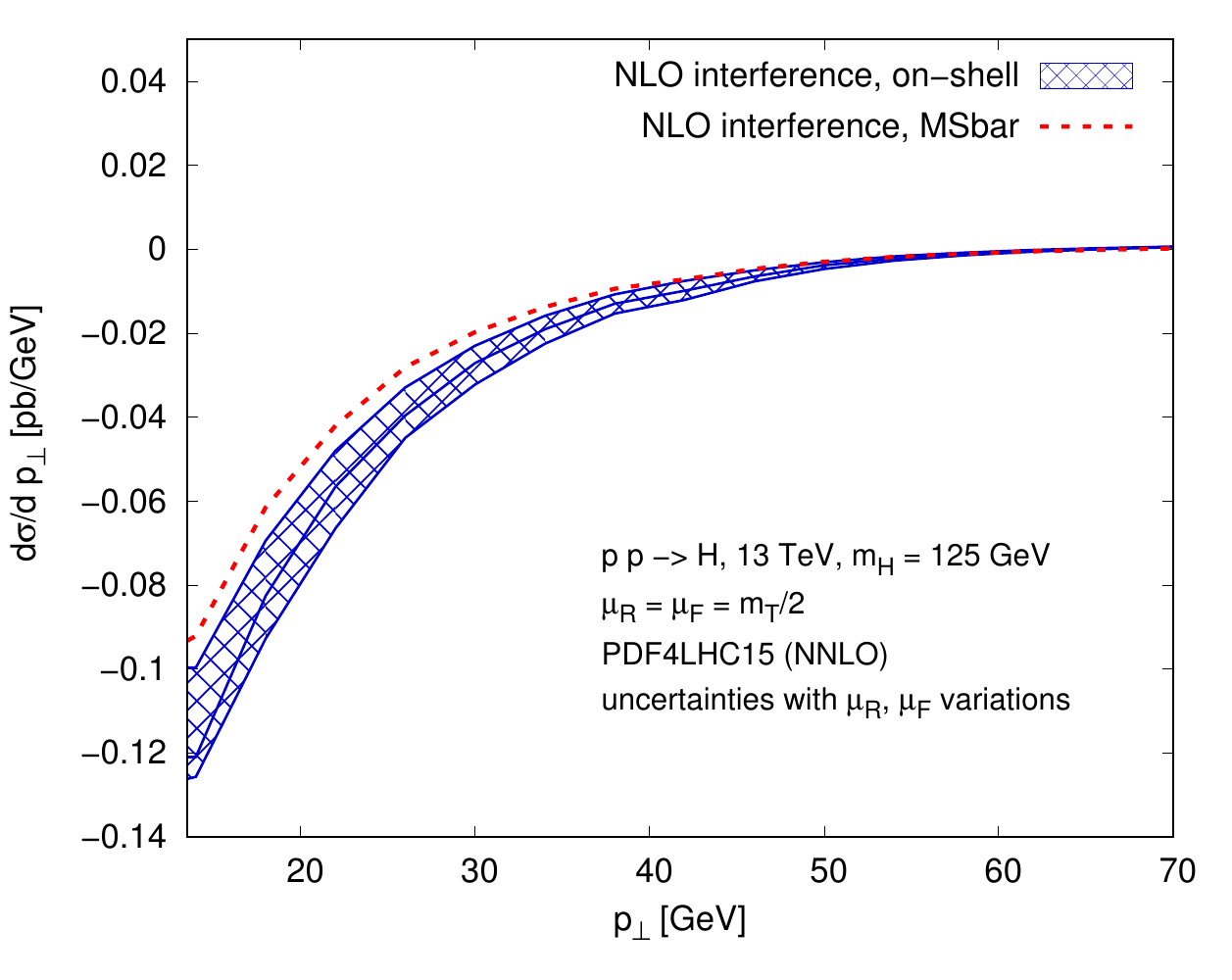}
\includegraphics[width=.48\textwidth]{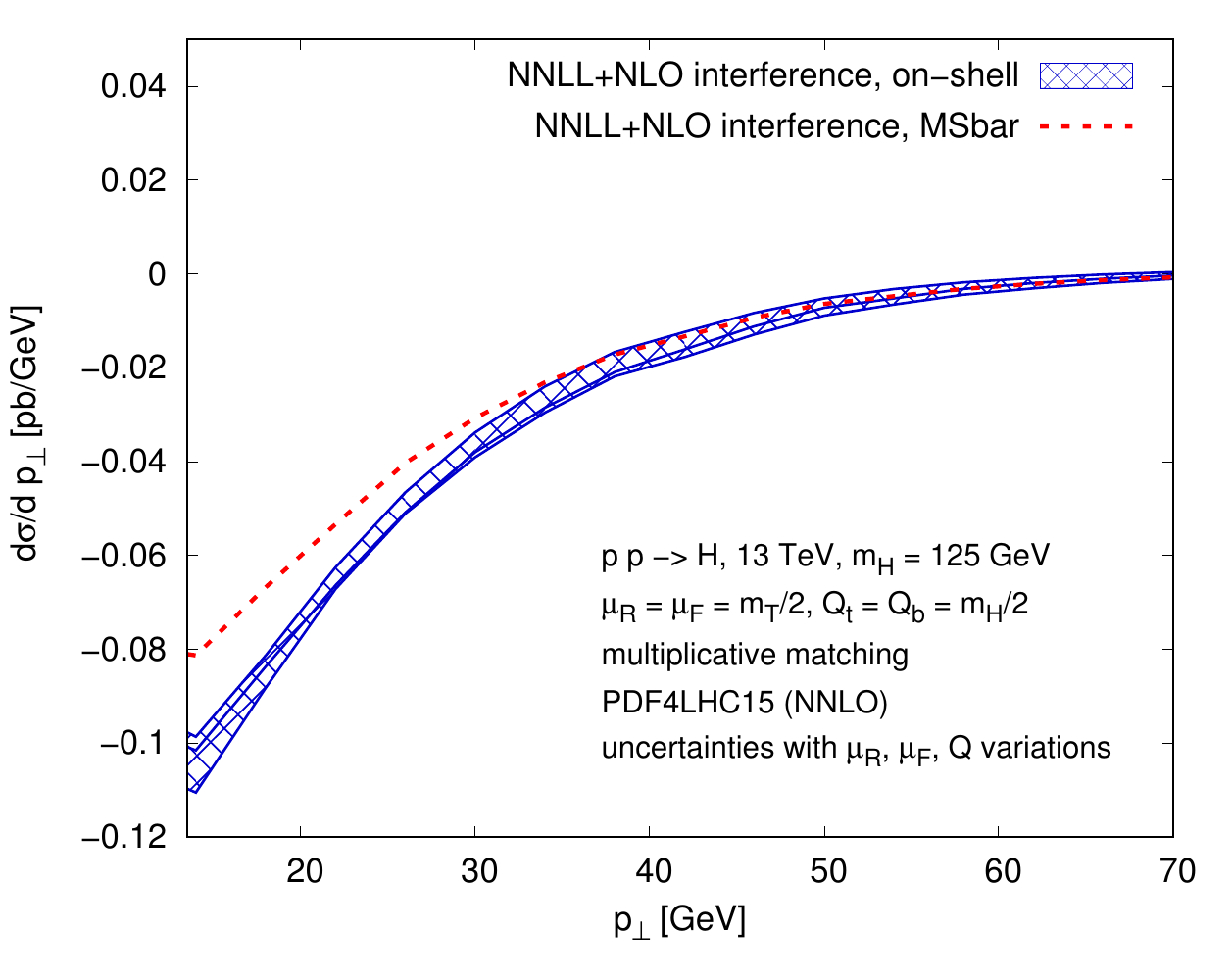}
\caption{The top-bottom interference contribution for the fixed order
  (left) and the matched (right) distributions, for different choices
  of the mass-renormalization scheme. See text for details.}
\label{fig:tb-mass-scheme}
\end{figure*}

In Fig.~\ref{fig:tb-mass-scheme} we display the results for the
top-bottom interference contribution. 
The fixed order result is presented in the left plot. 
We show the uncertainty band for the on-shell mass-renormalization 
scheme and the central value for the ${\overline{\rm MS}}$ scheme. 
The uncertainty band is calculated from a 7-point scale variation. 
We see that the scheme ambiguity is larger than the scale variation,
as already observed in~Ref.~\cite{Lindert:2017pky}.
The right plot shows
our results for the matched distributions with the two
different mass schemes. The difference between the two bottom-mass
schemes is similar to the fixed order case, but since now the
matched prediction has a smaller uncertainty, the separation between
the two results is more significant. 
As follows from Fig.~\ref{fig:tb-mass-scheme}, the
top-bottom interference contribution has an ambiguity of about
 $15-20\%$ down to $p_\perp \sim 10\, {\rm GeV}$. 
In order to improve on this, one would need a NNLO calculation
for the top-bottom interference, which is currently out of reach. 

\begin{figure*}[tb]
\centering
\hspace{-0.9cm} \includegraphics[width=.48\textwidth]{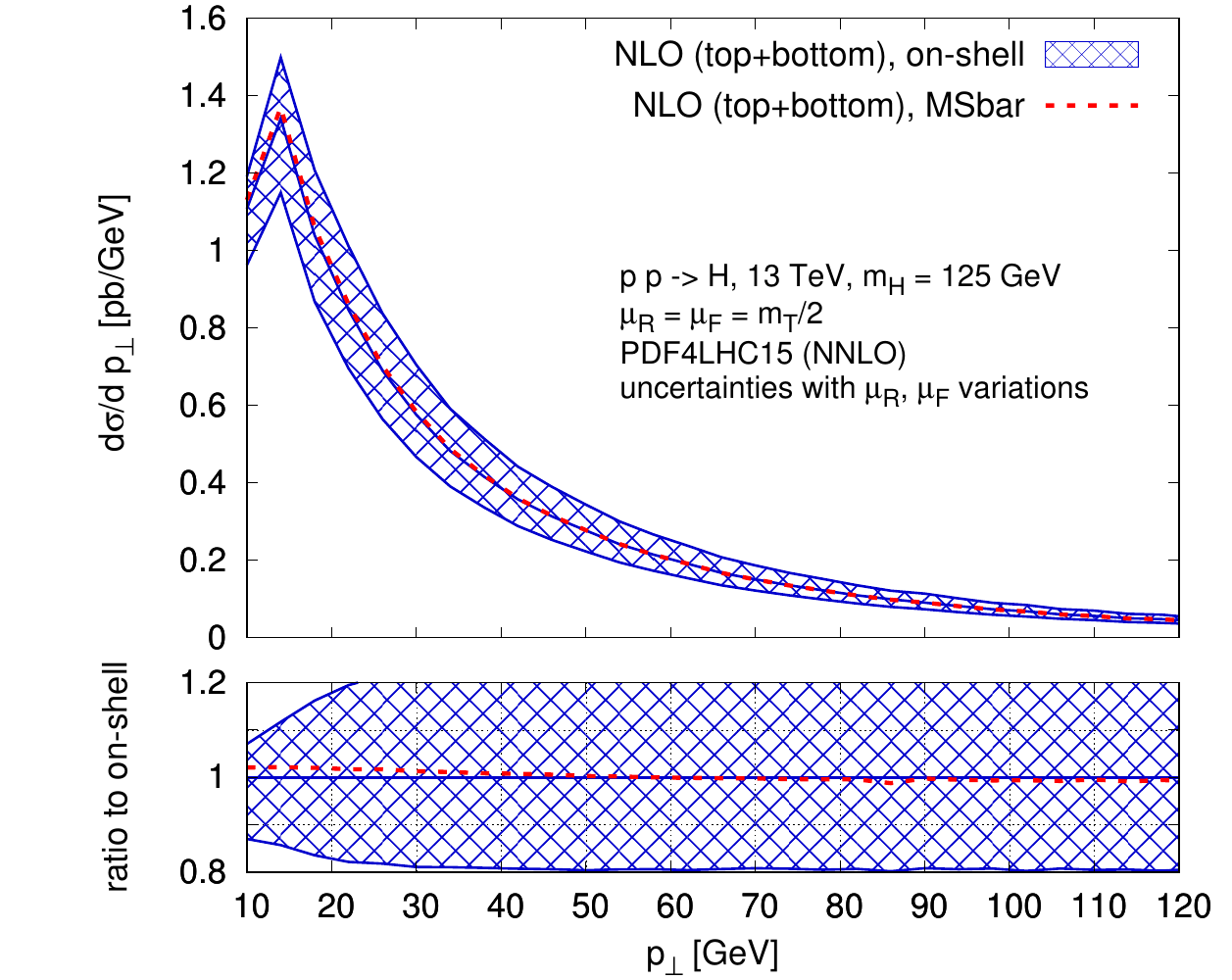} 
~~~
\hspace{-0.9cm} \includegraphics[width=.48\textwidth]{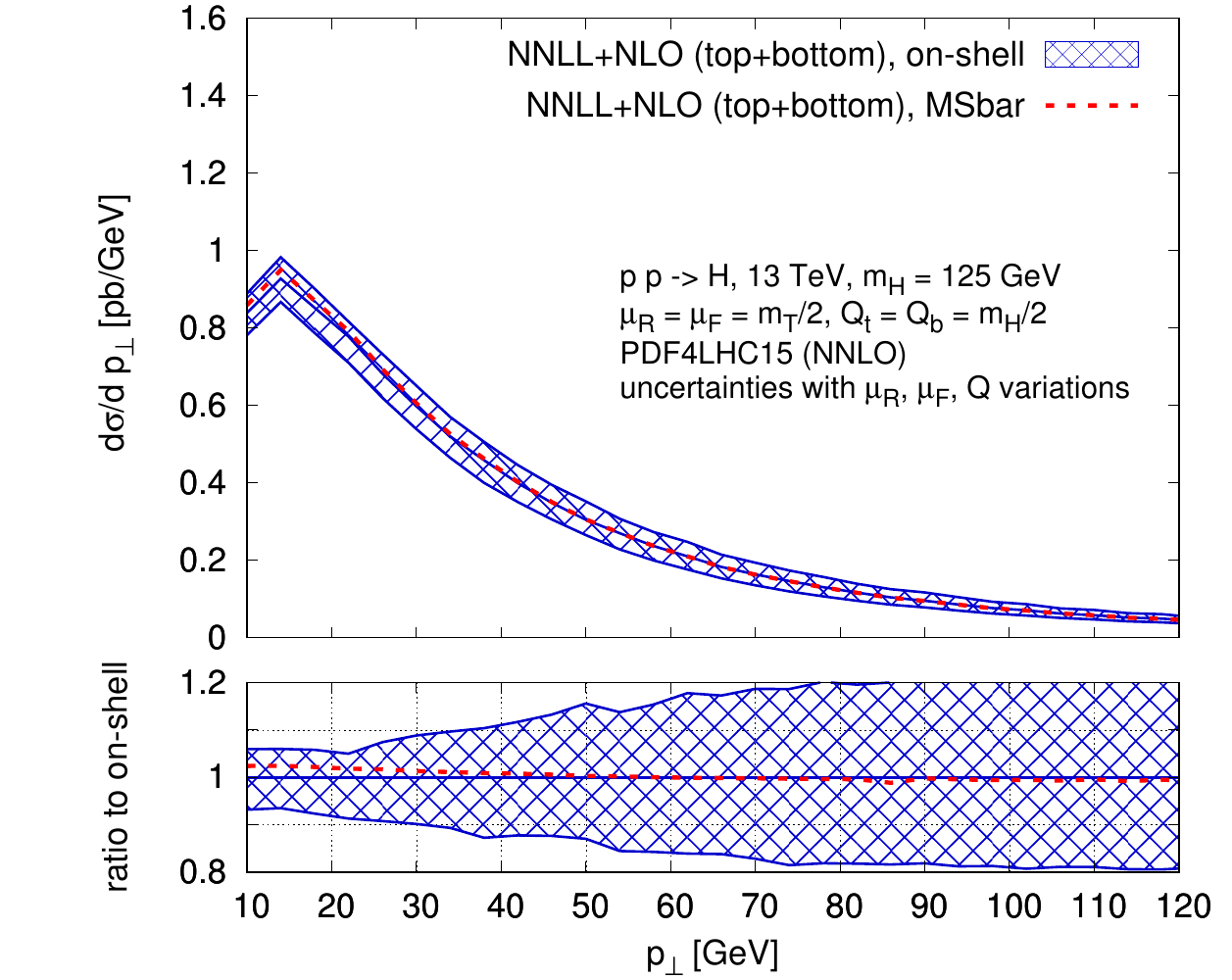} 
\caption{The full top plus bottom distribution for the fixed order (left) and the 
matched (right) results, for different choices of the mass renormalization scheme. See text for details.}
\label{fig:top-mass-scheme}
\end{figure*}

The analogous plots for the full distribution that includes both top
and bottom amplitudes are shown in Fig.~\ref{fig:top-mass-scheme}, for the
fixed order (left) and resummed (right) results. Unlike
for the top-bottom interference contribution, in this case the
difference between the two results for the bottom-mass schemes are
much smaller, at the level of a few percent. This is because
the top-bottom interference contributes to just $\mathcal O(-5\%)$ of the
full spectrum.\footnote{Although the bottom-mass scheme ambiguity has
  a very moderate impact on the SM Higgs $p_\perp$ spectrum, this
  effect might be more significant for specific BSM scenarios. A
  dedicated study of such scenarios is necessary in order to assess
  the theory uncertainties precisely.
} 

\begin{figure*}[htb]
\centering
\hspace{-0.1cm}\includegraphics[width=.5\textwidth]{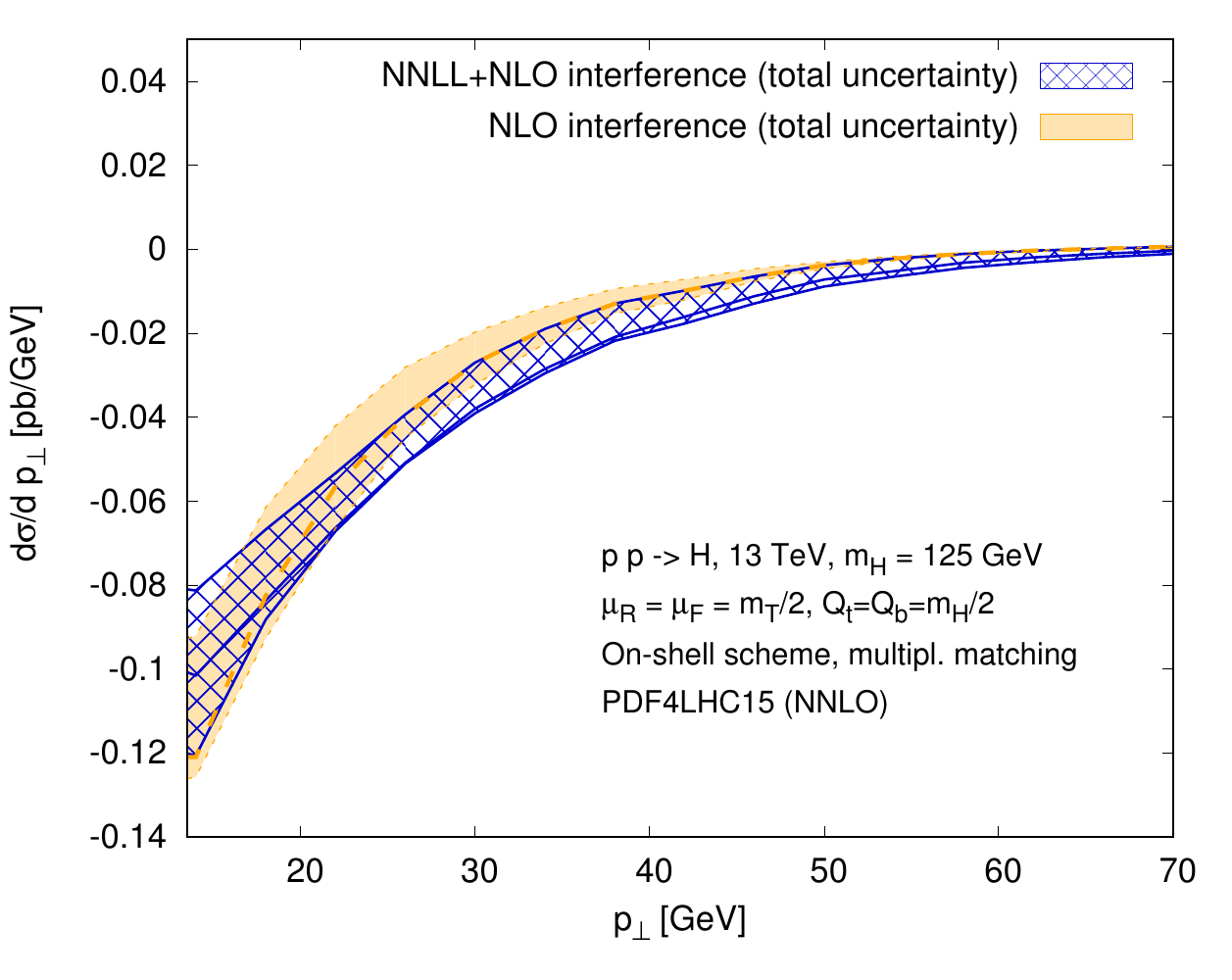} 
\hspace{-0.2cm}\includegraphics[width=.5\textwidth]{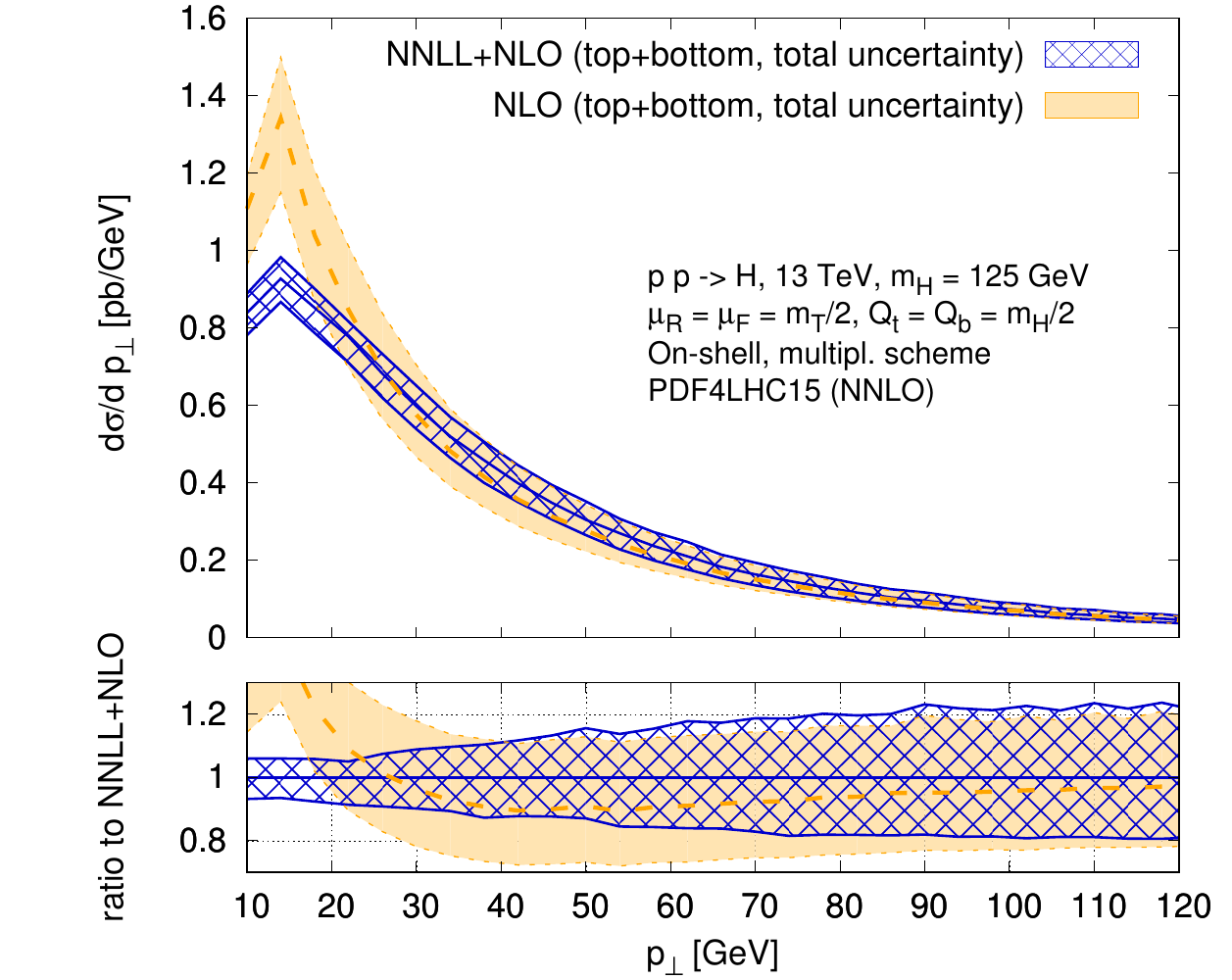}
\caption{The distributions for the top-bottom interference
  contribution (left) and the full NNLL matched result (right), using
  the multiplicative scheme with resummation scale $Q_b=Q_t=m_h/2$ as
  central values. See text for details.}
\label{fig:final-result}
\end{figure*}

Our best current predictions for the top-bottom interference and full
$\pt$ spectrum including all the relevant uncertainties are shown in
Fig.~\ref{fig:final-result}. As discussed earlier, the uncertainty bands 
are obtained as an envelope of:
\begin{itemize}
\item a 7-point renormalization and factorization scale variation;
\item resummation scale variation $Q_t=Q_b\in \{\mh/4,\mh/2,\mh\}$ for  $\mu_R=\mu_F=M_T/2$;
\item multiplicative matching scheme with $Q_t=\mh/2,~Q_b=2\mb$ for  $\mu_R=\mu_F=M_T/2$;
\item additive matching scheme with $Q_t=\mh/2,~Q_b=2\mb$ for  $\mu_R=\mu_F=M_T/2$.
\end{itemize}
In addition, if the default matched result but in the $\overline{\rm MS}$ renormalization scheme
for the bottom-quark mass is outside the uncertainty band estimated as described above,
we extend the uncertainty band to accommodate the mass scheme ambiguity. 
In fact, as shown in Figure~\ref{fig:tb-mass-scheme}, the latter ambiguity 
is the major source
of uncertainty  for the top-bottom interference for transverse momenta below $30~{\rm GeV}$. 

The top-bottom interference is shown in the left plot of
Fig.~\ref{fig:final-result}.  The qualitative features of the fixed-order 
result are unchanged by the resummation, which however has a
noticeable effect on the shape of the distribution. Our final result
has an uncertainty of about $\sim 20\%$, and is compatible with the
fixed-order one.  
In the right plot of Fig.~\ref{fig:final-result} we present the
results for the full spectrum.  
At large values
of the Higgs $p_{\perp} \gtrsim 30\, {\rm GeV}$ the fixed order result
is contained in the error band of the resummed result. However at
smaller values, $p_{\perp} \lesssim 30\, {\rm GeV}$ we observe a
marked difference between the two results. The error for the full
matched result is close to  $10\%$ for $\pt\lesssim 30~{\rm GeV}$
and close to $\sim 20\%$ at larger $p_{\perp}$. We stress however that
the uncertainty on the dominant top contribution can be further reduced by
employing the results of Refs.~\cite{Boughezal:2015aha,Boughezal:2015dra,Caola:2015wna,Chen:2014gva,Chen:2016zka,Monni:2016ktx,Bizon:2017rah}.

\section{Conclusions}
\label{sec:conclusions}

In this paper we performed a detailed study of the Higgs transverse 
momentum distribution, focusing on the region of intermediate values of 
transverse momenta, $\mb \lsim \pt \lesssim \mh$. Indeed, a precise 
theoretical control of the Higgs $\pt$ distribution in this region is 
essential to test the Higgs sector of the  Standard Model.   In particular, 
it provides a rare opportunity to probe the Yukawa couplings of 
light quarks, which are currently poorly constrained. In fact, although  
the main contribution to the Higgs production cross section is due to the  
coupling of the Higgs to top quarks, the coupling to bottom quarks has a 
non-negligible impact on the total cross section through its interference 
with the top,  decreasing the cross section  by about $\mathcal O(5\%)$.

The theoretical description of the Higgs $\pt$ distribution for $m_b\lsim \pt \lsim m_H$ 
in QCD is particularly challenging since, once
the contribution of bottom quarks is included, the
perturbative cross section for small $\pt$ suffers from the presence
of potentially large logarithms $\ln{\left( \pt / m_b\right)}$, $\ln{\left(
  m_H/m_b\right)}$, which can spoil the convergence of the perturbative expansion.  The
physical origin of these large logarithms is not yet fully understood,
and their all-order resummation remains currently out of reach.

Given these conceptual limitations, we provided our best theoretical
description of the Higgs $\pt$ distribution at NNLL+NLO QCD for
moderate values of the transverse momentum, including dependence
on the bottom mass. An important part of our study was a proper
assessment of the theory uncertainty of our results.
The NLO result for the top-bottom interference
suffers from scale uncertainties, which amount to around $15\%$. 
On top of this, a non-negligible source of uncertainty is
provided by the renormalization scheme ambiguity for the
bottom-quark mass, which we estimated by varying from the on-shell to
the $\MS$ scheme. This amounts to an uncertainty of up to $20
\%$ and it dominates the error budget of our prediction for the top-bottom
interference at small values of the Higgs $\pt$.  Together with the uncertainties associated with the
fixed order calculation, we also performed a detailed study of the
ones associated with the resummation procedure in the presence of bottom
quarks.  In order to estimate these ambiguities for the top-bottom
interference, we matched the fixed order NLO predictions with the NNLL
resummed cross-section using two different schemes, an additive
and a multiplicative one, and two very different choices of the
resummation scale, $Q_b=2m_b$ and $Q_b=m_H/2$.  
This leads to an uncertainty between $15-20\%$
on the top-bottom interference contribution to the $\pt$ spectrum.
Since the interference amounts to about $5\%$ of the full $\pt$ spectrum,
we conclude that unknown higher order $b$-quark mass effects can
modify the Higgs transverse momentum distribution by few percent.
All ambiguities associated with the resummation in the presence of
bottom quarks produce consistent results within the NNLL+NLO
uncertainty band, which is however driven by uncertainties in the
(NLO) top quark contribution. The latter is currently known to higher
N$^3$LL+NNLO
accuracy~\cite{Boughezal:2015aha,Boughezal:2015dra,Caola:2015wna,Chen:2014gva,
  Chen:2016zka,Monni:2016ktx,Bizon:2017rah}.
It
would be interesting to combine these results with the ones presented
in this article. We leave this for future investigations.

In conclusion, we presented a description of the Higgs $\pt$ spectrum
at NNLL+NLO QCD including both top and bottom quark contributions.
We found that the uncertainty on the top-bottom interference is $\mathcal O(20\%)$
in the region of interest $\mb\lesssim\pt\lesssim\mh$. 
Given the intrinsic ambiguities from scale
dependence and, in particular, from the choice of the bottom-mass
renormalization scheme and matching scheme, any improvement in this description will
inevitably require the computation of the NNLO QCD corrections to the
bottom-quark contribution to $gg \to H$ and $gg \to H+jet$.

\acknowledgments Some of us would like to thank the Munich
Institute for Astronomy and Particle Physics (MIAPP) for hospitality
and partial support during the programs \emph{Automated, Resummed and
  Effective} (F.C., K.M, P.F.M., L.T. and C.W.) and \emph{Mathematics
and Physics of Scattering Amplitudes} (F.C., L.T. and C.W.). 
The work of F.C. and L.T. was supported in part by the
ERC starting grant 637019 ``MathAm''.  The research of K.M. is
partially supported by BMBF grant 05H15VKCCA.  P.F.M. has been
supported by a Marie Sk\l{}odowska Curie Individual Fellowship of the
European Commission's Horizon 2020 Programme under contract number
702610 Resummation4PS.

\appendix

\section{Resummation and matching: details}
\label{sec:resummation}

In this appendix we briefly derive the resummation formula in the toy
model described in the main text, report the final NNLL formulas
that we eventually used in our results and describe in details
the matching procedure we employ. We follow the approach of
Refs.~\cite{Monni:2016ktx,Bizon:2017rah}, and we refer the reader to these publications for the
details.

\subsection{The LL case}
\label{app:LL}
We consider the $\pt$ distribution of a Higgs boson in $pp\to H$
in the $\pt\to 0$ limit, at leading-logarithmic accuracy. In this
approximation, one must only control the leading singularity of the
$n$-emissions matrix element at all perturbative orders. This is done
by approximating the process with an ensemble of independent
soft-collinear gluons emitted off the two incoming legs.

To set up the notation we introduce two reference light-like momenta along the beam direction that will
serve to parametrize the radiation\footnote{We remind the reader that
we are working in the soft approximation. As a consequence the
kinematics is much simplified.}
\begin{equation}
\label{eq:com-frame}
\tilde{p}_1 = \frac{\mh}{2}(1,0,0,1)\,,\qquad \tilde{p}_2 = \frac{\mh}{2}(1,0,0,-1) \,.
\end{equation}
We now consider a real emission $k_1$ collinear to $\tilde{p}_1$ that can be expressed as
\begin{equation}
  \label{eq:Sudakov}
  k_1 = \left(1-z_1^{(\ell_1)}\right)\tilde{p}_1 +  \left(1-z_1^{(\ell_2)}\right)
\tilde{p}_2 + \kappa_{\perp1}\,,
\end{equation}
where $\kappa_{\perp1}$ is a space-like four-vector, orthogonal to
both $\tilde{p}_1$ and $\tilde{p}_2$ such that
$\kappa_{\perp1}^2 = - k_{\perp1}^2$.  Note that since $k_1$
is massless
\[
k_{\perp1}^2 = \left(1-z_1^{(\ell_1)}\right)\left(1-z_1^{(\ell_2)}\right)\mhsq  
=\frac{2(\tilde{p}_1 \cdot k_1) 2(\tilde{p}_2 \cdot k_1)}{2(\tilde{p}_1 \cdot \tilde{p}_2)}\,.
\]
Moreover, if $k_1$ is collinear to $\tilde{p}_1$ one has (from $(1-z_1^{(\ell_1)}) >(1-z_1^{(\ell_2)})$)
\begin{equation}
\label{eq:zlimit}
z_1^{(\ell_1)} < 1-\frac{k_{\perp1}}{\mh}.
\end{equation}
An analogous limit on $z_1^{(\ell_2)}$ as in Eq.~\eqref{eq:zlimit} holds when $k_1$ is collinear to $\tilde{p}_2$.
Subsequent emissions off leg $\ell_1$ can be parametrized analogously to Eq.~\eqref{eq:Sudakov}, replacing the reference momentum $\tilde{p}_1 $ with
\begin{equation}
\tilde{p}_1 \to \left(\prod_{\substack{i=1\\\ell_i = \ell_1}}^k
  z_i^{(\ell_i)}\right)\tilde{p}_1\simeq \tilde{p}_1,
\end{equation}
where the product runs over all emissions off leg $\ell_1$ that occur
prior to the emission we are parametrizing, and we used the fact that
in the soft limit $ z_i^{(\ell_i)}\simeq 1$. A similar parametrization
holds for emissions off leg $\ell_2$.

The transverse recoil of the radiation
is absorbed entirely by the Higgs boson that acquires a transverse momentum
\begin{align}
\pt = |\sum_{i}\vec{k}_{\perp i}|.
\end{align}
In order to predict the $\pt\to 0$ limit, we need to sum emissions at all orders in the strong coupling.
With LL accuracy, the squared amplitude for $n$ emissions can be
approximated by a product of $n$ independent splitting kernels, as the
soft correlation between emissions starts contributing at NLL
order. The physical picture corresponding to this approximation is
given by a set of independent emissions off legs $\ell_1$ and
$\ell_2$. In this approximation, the differential partonic distribution
can be written as 
\begin{align}
\label{eq:eikonal_app}
\frac{{\rm d}\hat \sigma}{{\rm d}\pt} \simeq 
[d p_H]\mathcal|\mathcal M(\tilde p_1 + \tilde p_2\to
H)|^2 \delta^{(4)}(\tilde p_1+\tilde p_2 - p_H)\notag
\\
\times \frac{1}{n!}\prod_{i=1}^{n}  [dk_i]|M_{\rm soft}(k_i)|^2
\delta\left(\pt- |\sum_{i}\vec{k}_{\perp i}|\right),
\end{align}
where the eikonal squared amplitude for a single emission reads
\begin{align}
  \label{eq:single-emsn}
  [dk] | M_{\rm soft}(k)|^2=\sum_{\ell=1,2} 2 C_A \frac{\alpha_s(k_{\perp})}{\pi}\frac{dk_{\perp}}{k_{\perp}} 
 \frac{d z^{(\ell)}}{1-z^{(\ell)}}\, \Theta\left((1-z^{(\ell)}) - k_{\perp}/\mh\right) 
\frac{d\phi}{2\pi}.
\end{align}
In Eq.~\eqref{eq:single-emsn}, the coupling is evaluated at $k_{\perp}$ to account for the
leading-logarithmic contribution of the gluon branching into either a
pair of soft quarks or gluons, see e.g.~\cite{Banfi:2004yd} for a detailed
explanation.

The resummation is naturally performed at the level of the cumulative distribution, defined as
\begin{equation}
\Sigma(\pt) = \int_0^{\pt} d \pt' \frac{d\sigma}{d \pt'}.
\end{equation}
Indeed while the differential spectrum involves plus distributions in
$\pt$, $\Sigma(\pt)$ is a regular function. 
From Eq.~\eqref{eq:eikonal_app}, it follows that the cumulative distribution with LL accuracy 
can be written as
\begin{equation}
\Sigma(\pt) \simeq \left[f_g(\mu_F)\otimes f_g(\mu_F)\right](\mh^2/s)\times\int d\hat\sigma\,\Theta\left(\pt- |\sum_{i}\vec{k}_{\perp i}|\right),
\end{equation}
where $f_g(\mu_F)$ is the gluon parton density evaluated at the factorization scale $\mu_F$, and
the convolution is defined as usual
\begin{equation}
\left[f\otimes g\right](x) \equiv \int_0^1 {\rm d}y \; {\rm d}z \; \delta(x-yz) f(y) g(z).
\end{equation}
Since $\pt$ only constrains the transverse momentum of the emissions,
we can perform the integrals over the $z_i^{(\ell_i)}$ components
inclusively. It is therefore convenient to introduce the functions
\begin{equation}
  \label{eq:R'1,2}
  \begin{split}
    R'_1\left(\pt\right)&= \int [dk]
    |M_{\rm soft}(k)|^2
    \,(2\pi) \delta(\phi-\bar \phi)\, \pt\delta\left(\pt-k_{\perp}\right)\Theta(z^{(2)}
-z^{(1)})  \,, \\
    R'_2\left(\pt\right)& = \int [dk]
    |M_{\rm soft}(k)|^2\,(2\pi) \delta(\phi-\bar \phi)\,
    \pt\delta\left(\pt-k_{\perp}\right)\Theta(z^{(1)}
-z^{(2)})\,.
  \end{split}
\end{equation}
This notation allows us to parametrize the real-emission matrix
element and phase space as
\begin{align}
  \label{eq:single-emsn-rp}
    [dk_i] |M_{\rm soft}(k_i)|^2  = \frac{dk_{\perp i}}{k_{\perp i}}
    \frac{d\phi_i}{2\pi}\sum_{\ell_i=1,2}
    R'_{\ell_i}\left(k_{\perp i}\right)
    &= \frac{d\zeta_i}{\zeta_i} \frac{d\phi_i}{2\pi} \sum_{\ell_i=1,2}
    R'_{\ell_i}\left(\zeta_i k_{\perp1}\right) \,,
\end{align}
where we defined $\zeta_i = k_{\perp i}/k_{\perp1}$. 

We now discuss the purely virtual corrections, which are encoded in
the gluon form factor $|\mathcal M(\tilde p_1 + \tilde p_2\to
H)|^2$. We write it as
\begin{equation}
|\mathcal M(\tilde p_1 + \tilde p_2\to
H)|^2 = {\cal H}(\mh)|\mathcal M_B(\tilde p_1 + \tilde p_2\to
H)|^2,
\end{equation}
where the function ${\cal H}$ contains all the IRC singularities and
the constant finite corrections of the form factor, and ${\cal M}_B$
denotes the Born amplitude. Since we are
working with LL accuracy, we are only interested in the leading
singular term of ${\cal H}$ at all orders (while neglecting all finite
terms) which can be written as 
\begin{equation}
\label{eq:form_factor}
{\cal H}(\mh) \simeq \exp\left\{-\int [dk] |M_{\rm soft}(k)|^2\right\}.
\end{equation}
Note that the integral in Eq.~\eqref{eq:form_factor} is divergent and is to be
considered as regularized. In order to cancel the IRC divergences of
the real emissions~\eqref{eq:eikonal_app} against the ones in the virtual
corrections~\eqref{eq:form_factor} at all orders, we introduce a
small slicing parameter $\epsilon >0$ such that all emissions with a
transverse momentum $k_{\perp i}$ smaller than $\epsilon k_{\perp1}$ can be
ignored in the computation of the observable $\pt$, in the limit
$\epsilon\to 0$. The real emissions with $k_{\perp i}<\epsilon k_{\perp1}$,
hereby denoted as {\it unresolved}, can be directly combined with the
virtual corrections at all orders. Their combination gives rise to an exponential
suppression factor of the type
\begin{align}
{\cal H}(\mh)&\sum_{m}^{\infty}\frac{1}{m!}\prod_{i=1}^m \bigg[\int\frac{dk_{\perp i}}{k_{\perp i}}
    \frac{d\phi_i}{2\pi}\sum_{\ell_i=1,2}
    R'_{\ell_i}\left(k_{\perp i}\right) \Theta\left(\epsilon
      k_{\perp1}-k_{\perp i}\right) \bigg] \notag\\
&= \exp\left\{ -\int\frac{dk_{\perp}}{k_{\perp}}
    \frac{d\phi}{2\pi}\sum_{\ell=1,2}
    R'_{\ell}\left(k_{\perp}\right) \Theta\left(k_{\perp} - \epsilon
      k_{\perp1}\right)\right\}\equiv e^{-R(\epsilon k_{\perp1})}.
\end{align}

On the other end, emissions with $k_{\perp i}>\epsilon k_{\perp1}$, that we
denote as {\it resolved}, are constrained by the
observable's measurement function and therefore cannot be integrated
over inclusively. The resummed LL cross section thus reads
\begin{align}
\label{eq:XS_LL}
\Sigma(\pt) \simeq &\,\sigma_B \int \frac{dk_{\perp1}}{k_{\perp1}}
    \frac{d\phi_1}{2\pi} e^{-R(\epsilon k_{\perp1})}\sum_{\ell_1=1,2}
    R'_{\ell_1}\left(k_{\perp1}\right)\notag\\
&\times  \sum_{n=0}^{\infty}\frac{1}{n!}\prod_{i=2}^{n+1}\int_{\epsilon}^1 \frac{d\zeta_i}{\zeta_i} \frac{d\phi_i}{2\pi} \sum_{\ell_i=1,2}
    R'_{\ell_i}\left(\zeta_i k_{\perp1}\right) \Theta\left(\pt- |\sum_{i}\vec{k}_{\perp i}|\right),
\end{align}
where $\sigma_B$ is the Born cross section. The above formula, in the
limit $\epsilon\to 0$ exactly reproduces the LL corrections to the
$\pt$ distribution, see Ref.~\cite{Bizon:2017rah} for a formal
proof. Eq~\eqref{eq:XS_LL} can be further simplified by observing that
in the resolved radiation one always has $\zeta_i\sim 1$, since
configurations in which $\zeta_i\ll 1$ are automatically canceled
against the exponential Sudakov factor $e^{-R(\epsilon
  k_{\perp1})}$. Therefore, one can expand the functions
$R'\left(\zeta_i k_{\perp1}\right)$ in powers of $\ln(1/\zeta_i)$ as
\begin{align}
\label{eq:expansion}
R'_{\ell_i}\left(\zeta_i  k_{\perp1}\right) = R'_{\ell_i}\left(k_{\perp1}\right)
  +R''_{\ell_i}\left(k_{\perp1}\right) \ln\frac{1}{\zeta_i} +  \dots,
\end{align}
and retain terms that contribute at a given logarithmic order. In
particular, at LL, only the first term in this expansion contributes,
and higher-order terms matter at higher logarithmic orders (see
Refs.~\cite{Bizon:2017rah} for details).

Similarly, we can consistently expand out the $\epsilon$
dependence of the exponential Sudakov as
\begin{equation}
e^{-R(\epsilon k_{\perp1})}= e^{-R(k_{\perp1})} e^{-R'\left(k_{\perp1}\right)
  \ln\frac{1}{\epsilon} + \dots },
\end{equation}
where the $\epsilon$ dependence manifestly cancels against the one in
the resolved contribution, and we defined
\begin{equation}
R'\left(k_{\perp1}\right)\equiv \sum_{\ell_1=1,2}
    R'_{\ell_1}\left(k_{\perp1}\right).
\end{equation}
Therefore, with LL accuracy,
Eq.~\eqref{eq:XS_LL} becomes
\begin{align}
\label{eq:XS_LL_expanded}
\Sigma(\pt) \simeq &\,\sigma_B \int \frac{dk_{\perp1}}{k_{\perp1}}
    \frac{d\phi_1}{2\pi} e^{-R(k_{\perp1})}\epsilon^{R'\left(k_{\perp1}\right)}\sum_{\ell_1=1,2}
    R'_{\ell_1}\left(k_{\perp1}\right)\notag\\
&\times  \sum_{n=0}^{\infty}\frac{1}{n!}\prod_{i=2}^{n+1}\int_{\epsilon}^1 \frac{d\zeta_i}{\zeta_i} \frac{d\phi_i}{2\pi} \sum_{\ell_i=1,2}
    R'_{\ell_i}\left(k_{\perp1}\right) \Theta\left(\pt- |\sum_{i}\vec{k}_{\perp i}|\right).
\end{align}
Equation~\eqref{eq:XS_LL_expanded} is suitable for a numerical
implementation, as explained in Ref.~\cite{Bizon:2017rah} in
detail. The dependence on $\epsilon$ is at most power suppressed
(i.e. ${\cal O}(\epsilon \pt)$) and it vanishes in the limit $\epsilon\to
  0$. This limit can therefore be taken safely numerically, and the
  result is absolutely stable for very small values of
  $\epsilon$.\footnote{In our implementation we use $\epsilon =
    e^{-20}$, although any value below $\epsilon =
    e^{-6}$ does not lead to any appreciable differences.}

We now introduce the resummation scale $Q$ as a possible way to switch
off the resummation at large transverse momentum. This is defined with
a procedure similar to the one discussed in the text. We first break the
logarithm as follows
\be
L\equiv\ln \frac{\mh}{k_{\perp1}} = \ln\frac{\mh}{Q}+\ln\frac{Q}{k_{\perp1}}.
\ee
The above operation will allow us (as explained shortly) to have an
additional handle (namely the scale $Q$) to estimate the size of
subleading logarithmic terms. Moreover, we also slightly modify the
phase space available for the radiation, by introducing
power-suppressed contributions that ensure that at large $\pt$ the
resummation effects completely vanish. This can be done, as a first
step, by modifying the resummed logarithms as follows
\be
\ln\frac{Q}{k_{\perp1}} \to \frac{1}{p}\ln \left(\frac{Q^p}{k_{\perp1}^p}+1\right) \equiv \tilde L,
\label{eq:modlog_kt1}
\ee
where $p$ is a positive real parameter which is chosen such that the
resummed differential distribution vanishes as $1/\pt^{p+1}$ at large
$\pt$. The above prescription essentially amounts to the
following
\begin{enumerate}
\item First, we split the
resummed logarithm $L$ into the sum of a {\it small} logarithm
$\ln(\mh/Q)$ (with $Q\sim \mh$) and a {\it large} one
$\ln(Q/k_{\perp1})$. This operation allows one to introduce a generic scale
$Q$ which appears in the resummed logarithms. One can now expand $L$
about $\ln(Q/k_{\perp1})$, retaining all terms with the desired logarithmic
accuracy. Effectively, this implies that $\ln(\mh/Q)$ is treated
perturbatively at fixed order. Moreover, we replace $\ln(Q/k_{\perp1})$ by
the modified logarithm $\tilde{L}$. In our LL example this means
\begin{align}
R(k_{\perp1}) \to \tilde{R}(k_{\perp1}) + {\cal O}(\ln \mh/Q);\qquad
R(k_{\perp1}) \to \tilde{R}'(k_{\perp1}) + {\cal O}(\ln \mh/Q),
\end{align}
where $\tilde{R}$ and $\tilde{R}'$ are functions of the modified
logarithm $\tilde{L}$ only.
\item Eq.~\eqref{eq:modlog_kt1} comes together with the following
  prefactor $\mathcal J$ in Eq.~\eqref{eq:XS_LL_expanded}
\begin{equation}
\label{eq:jakobian}
{\cal J}(k_{\perp1}) = \left(\frac{Q}{k_{\perp1}} \right)^p \left(1+\left(\frac{Q}{k_{\perp1}} \right)^p\right)^{-1}.
\end{equation} 
This corresponds to the Jacobian for the
transformation~\eqref{eq:modlog_kt1}, and ensures the absence of
fractional (although power suppressed) $\alpha_s$ powers in the final
distribution~\cite{Bizon:2017rah}.  This factor, once again, leaves
the small $k_{\perp1}$ region untouched, and only modifies the large
$\pt$ region by power-suppressed effects. This is effectively mapping
the limit $k_{\perp1}\to Q$ onto $k_{\perp1}\to \infty$. Although this
procedure seems a simple change of variables, we stress that the
observable's measurement function (i.e. the $\Theta$ function in
Eq.~\eqref{eq:XS_LL_expanded}) is not affected by this
prescription. As a consequence, the final result will depend on the
parameter $p$ through power-suppressed terms.
\end{enumerate}

The difference between the above prescription and what was introduced
in the text is that the argument of the (modified) logarithms is now
$k_{\perp1}$ instead of $\pt$. This prescription is technically more
correct, since in the small $k_{\perp1}$ region, which governs the $\pt\to
0$ limit, the modified logarithms leave Eq.~\eqref{eq:XS_LL_expanded}
untouched. Conversely, at large $k_{\perp1}$, where one has $k_{\perp1}\sim
\pt$, the above prescription reduces to what was defined in the text,
i.e. the modified logarithms of $k_{\perp1}$ in this region are formally
equivalent to modified logarithms in $\pt$. To see this, we observe
that when $k_{\perp1}\gg Q$ the function $R'(k_{\perp1})\ll 1$. Therefore, the
probability of having any emission after the first one in
Eq.~\eqref{eq:XS_LL_expanded} is strongly suppressed. As a
consequence, at large $k_{\perp1}$, the only relevant event is the one
that involves a single emission $k_1$, for which the cross section
reads
\begin{align}
\label{eq:XS_LL_asympt}
\Sigma(\pt) \sim &\,\sigma_B \int \frac{dk_{\perp1}}{k_{\perp1}}
    \frac{d\phi_1}{2\pi} {\cal J}(k_{\perp1}) e^{-\tilde{R}(k_{\perp1})}\sum_{\ell_1=1,2}
    \tilde{R}'_{\ell_1}\left(k_{\perp1}\right)\Theta\left(\pt-
                   |\vec{k}_{\perp1}|\right) = e^{-\tilde{R}(\pt)}.
\end{align}
It is easy to see that, if Eq.~\eqref{eq:XS_LL_asympt} were evaluated
without the factor ${\cal J}$, it would lead to additional
power-suppressed terms with fractional power of the coupling, which
are clearly spurious.

\subsection{Final formulas for NNLL resummation}
\label{sec:NNLL}
Beyond LL, Eq.~\eqref{eq:XS_LL_expanded} is corrected to account for
the description of the real-emission matrix element and phase space in
less singular configurations, as well as higher perturbative
corrections. To NNLL order it can be expressed as~\cite{Bizon:2017rah}
\begin{align}
\label{eq:NNLL}
\Sigma(\pt) &= \int\frac{d k_{\perp1}}{k_{\perp1}} 
               \frac{d\phi_1}{2\pi} {\cal J}(k_{\perp1})\left[\epsilon^{{
                                                        \tilde R}'(k_{\perp1})}\sum_{n=0}^\infty\frac{1}{n!}\prod_{i=2}^{n+1}\int_{\epsilon}^1\frac{d\zeta_i}{\zeta_i}\frac{d\phi_i}{2\pi}\sum_{\ell_1=1,2}
    \tilde R'_{\ell_1}\left(k_{\perp1}\right)\right]\notag\\
&\times
  \Bigg\{\frac{d}{d \tilde L}\left[-e^{-\tilde R(k_{\perp1})}{\cal
  L}(k_{\perp1})\right]\Theta\left(\pt-|\sum_{i=1}^{n+1}\vec{k}_{\perp i}|
  \right)
+ e^{-\tilde R(k_{\perp1})}{ \tilde R}'(k_{\perp1})
\notag\\
& \times\int_{\epsilon}^1\frac{d \zeta_{s}}{\zeta_{s}} 
                     \frac{d\phi_s}{2\pi} \left[\sum_{\ell_i=1,2}\left(\delta{
                                                        \tilde R}'_{\ell_i}(k_{\perp1}) + {
                                                        \tilde R}''_{\ell_i}(k_{\perp1})\ln\frac{k_{\perp1}}{k_{\perp s}}\right)
  {\cal{\hat L}}(k_{\perp1}) -  \frac{d {\cal {\hat L}}(k_{\perp1})}{d
 \tilde L}\right]  \notag\\
&\times\bigg[\Theta\big(\pt-|\sum_{i=1}^{n+2}\vec{k}_{\perp i}|
  \big)-\Theta\big(\pt-|\sum_{\substack{i=1\\ i \neq s}}^{n+2}\vec{k}_{\perp i}|
  \big)\bigg]\Bigg\},
\end{align}
where $\zeta_i=k_{\perp i}/k_{\perp1}$.
In this formula, the Sudakov radiator
$\tilde R(k_{\perp1})$ is corrected with respect to its LL expression by
higher-order corrections of both soft and collinear origin. The same
comment applies to the function $\tilde R'_{\ell_i}$ which we decided to
split into the old $\tilde R'$ (derivative of the LL radiator defined above),
plus a correction that contains all subleading effects, therefore
replacing $\tilde R'$ with
\begin{equation}
\tilde R'_{\ell_i} \to \tilde R'_{\ell_i} + \delta \tilde R'_{\ell_i}.
\end{equation}
The correction due to $\delta \tilde R'_{\ell_i}$ is only relevant to NNLL
order for one of the resolved emissions. This {\it special} emission
is denoted by the subscript $s$ in Eq.~\eqref{eq:NNLL}. After
expanding to first order the corresponding term proportional to $\delta
\tilde R'_{\ell_i}$ arising from the initial $\epsilon^{\tilde R'}$ factor, one ends
up with the second term in the curly bracket in
Eq.~\eqref{eq:NNLL}, see Ref.~\cite{Bizon:2017rah} for a full derivation. 
The same manipulations apply to the $\tilde R''$
correction coming from the expansion~\eqref{eq:expansion} discussed
above.

Moreover, we introduced the following generalized luminosity coefficient
\begin{align}
\label{eq:luminosity-NNLL0}
&{\cal L}(k_{\perp1}) = |{\cal M}_{B}|^2 \sum_{i, j}\int d x_1 d
  x_2\int_{x_1}^{1}\frac{d z_1}{z_1}\int_{x_2}^{1}\frac{d
  z_2}{z_2}f_i\!\left(\mu_F e^{-\tilde L},\frac{x_1}{z_1}\right)f_{j}\!\left(\mu_F e^{-\tilde L},\frac{x_2}{z_2}\right)\delta(x_1
  x_2 s - \mh^2)\notag\\
&\times\Bigg[\delta_{gi}\delta_{gj}\delta(1-z_1)\delta(1-z_2)
\left(1+\frac{\alpha_s(\mu_R)}{2\pi} {\tilde H}^{(1)}\left(\mu_R,\frac{Q}{\mh}\right)\right) \notag\\
&+ \frac{\alpha_s(\mu_R)}{2\pi}\frac{1}{1-2\alpha_s(\mu_R)\beta_0
  \tilde L}\left({\tilde C}_{g i}^{(1)}\left(z_1,\mu_F,\frac{Q}{\mh}\right)\delta(1-z_2)\delta_{gj}+
  \{z_1\leftrightarrow z_2; i \leftrightarrow j\}\right)\Bigg],
\end{align}
and its NLL approximation
\begin{align} 
\label{eq:luminosity-NLL0}
{\hat {\cal L}}(k_{\perp1})  =|{\cal M}_{B}|^2 \int d x_1 d
  x_2 f_g\!\left(\mu_F e^{-\tilde L},x_1\right)f_{g}\!\left(\mu_F e^{-\tilde L},x_2\right) \delta(x_1
  x_2 s - \mh^2).
\end{align}

We now report all the various ingredients entering the above formulas.
The ${\cal O}(\alpha_s)$ correction to the collinear coefficient
functions reads
\begin{align}
 {\tilde C}_{g i}^{(1)}\left(z,\mu_F,\frac{Q}{\mh}\right)&= -P_{ij}^{(0),\epsilon}(z)-\delta_{ij}\delta(1-z)C_A \frac{\pi^2}{12}+P_{ij}^{(0)}(z)
  \ln{\frac{Q^2}{\mu_{F}^{2}}}
\end{align}
where $P_{ij}^{(0)}$ are the LO Altarelli-Parisi splitting functions
\begin{subequations}
\begin{align}
\label{eq:regAP}
 P^{(0)}_{qq}(z)&=C_F\left[\frac{1+z^2}{(1-z)_+}+\frac32\delta(1-z)\right],\nonumber\\
 P^{(0)}_{qg}(z)&=T_R\left[z^2+(1-z)^2\right],\nonumber\\
 P^{(0)}_{gq}(z)&=C_F\frac{1+(1-z)^2}{z},\nonumber\\
 P^{(0)}_{gg}(z)&=2C_A\left[\frac z{(1-z)_+}+\frac{1-z}z+z(1-z)\right]+2\pi\beta_0\delta(1-z),
\end{align}
\end{subequations}
with $\beta_0 = (11 C_A - 2 n_f)/(12\pi)$ 
and $P_{ij}^{(0),\epsilon}(z)$ are given by
\begin{subequations}
\begin{align}
 P_{qq}^{(0),\epsilon}(z) &= -C_F(1-z)\,,\\
 P_{gq}^{(0),\epsilon}(z) &= -C_F z\,,\\
 P_{qg}^{(0),\epsilon}(z) &= -2 T_R z(1-z)\,,\\
 P_{gg}^{(0),\epsilon}(z) &= 0.
\end{align}
\end{subequations}
The function $ {\tilde
  H}^{(1)}\left(\mu_R,\frac{Q}{\mh}\right)$ is defined as
\begin{align}
  {\tilde H}^{(1)}\left(\mu_R,\frac{Q}{\mh}\right)&= H^{(1)}- \left( B^{(1)}+\frac{A^{(1)}}{2}\ln{\frac{\mh^2}{Q^2}}\right)\ln{\frac{\mh^2}{Q^2}}
 + 4\pi\beta_{0}\ln{\frac{\mu_{R}^2}{\mh^{2}}}\,,
\end{align}
where $H^{(1)}$ denotes the finite one-loop virtual correction to the $gg\to
H$ process and $A^{(i)},B^{(i)}$ are reported below. For the top contribution in the
$m_t\to\infty$ approximation, $H^{(1)}$ reads
\begin{equation}
H^{(1)} = C_A\left(5+\frac{7}{6}\pi^2\right)-3C_F = 11 + \frac{7}{2}\pi^2\,.
\end{equation}
The result including full quark mass dependence has been computed 
analytically in Refs.~\cite{Harlander:2005rq,Aglietti:2006tp,Anastasiou:2006hc}.\footnote{In our implementation
we take both the Born amplitude and the virtual corrections from Ref.~\cite{Aglietti:2006tp}.}

We expand the Sudakov radiator as
\begin{equation}
\tilde{R} = \tilde{L} g_{1}(\lambda) + g_{2}(\lambda) + \frac{\alpha_s(\mh)}{\pi}g_{3}(\lambda),
\end{equation}
where 
\begin{equation}
\lambda = \as(\mu_R) \beta_0 \tilde L.
\end{equation}
We introduce 
\begin{equation}
x_Q = \frac{Q}{\mh},
\end{equation}
and write
\begin{align}
  g_{1}(\lambda) =& \frac{A^{(1)}}{\pi\beta_{0}}\frac{2 \lambda +\ln (1-2 \lambda )}{2  \lambda }, \\
  g_{2}(\lambda) =& \frac{1}{2\pi \beta_{0}}\ln (1-2 \lambda )
  \left(A^{(1)} \ln \frac{1}{x_Q^2}+B^{(1)}\right)
  -\frac{A^{(2)}}{4 \pi ^2 \beta_{0}^2}\frac{2 \lambda +(1-2
    \lambda ) \ln (1-2 \lambda )}{1-2
    \lambda} \notag\\
  &+A^{(1)} \bigg(-\frac{\beta_{1}}{4 \pi \beta_{0}^3}\frac{\ln
    (1-2 \lambda ) ((2 \lambda -1) \ln (1-2 \lambda )-2)-4
    \lambda}{1-2 \lambda}\notag\\
  &\hspace{10mm}-\frac{1}{2 \pi \beta_{0}}\frac{(2 \lambda(1
    -\ln (1-2 \lambda ))+\ln (1-2 \lambda ))}{1-2\lambda} \ln
  \frac{\mu_R^2}{x_Q^2 \mh^2}\bigg)\,,\\
  g_{3}(\lambda) =
  & \left(A^{(1)} \ln\frac{1}{x_Q^2}+B^{(1)}\right)
  \bigg(-\frac{\lambda }{1-2 \lambda} \ln
  \frac{\mu _{R}^2}{x_Q^2\mh^2}+\frac{\beta_{1}}{2 \beta_{0}^2}\frac{2 \lambda
    +\ln (1-2 \lambda )}{1-2 \lambda}\bigg)\notag\\
  &   -\frac{1}{2 \pi\beta_{0}}\frac{\lambda}{1-2\lambda}\left(A^{(2)}
  \ln\frac{1}{x_Q^2}+B^{(2)}\right)-\frac{A^{(3)}}{4 \pi ^2 \beta_{0}^2}\frac{\lambda ^2}{(1-2\lambda )^2} \notag\\
  &   +A^{(2)} \bigg(\frac{\beta_{1}}{4 \pi  \beta_{0}^3 }\frac{2 \lambda  (3
    \lambda -1)+(4 \lambda -1) \ln (1-2 \lambda )}{(1-2 \lambda
    )^2}-\frac{1}{\pi \beta_{0}}\frac{\lambda ^2 }{(1-2 \lambda )^2}\ln\frac{\mu_R^2}{x_Q^2 \mh^2}\bigg) \notag\\
  & +A^{(1)} \bigg(\frac{\lambda  \left(\beta_{0} \beta_{2} (1-3 \lambda
    )+\beta_{1}^2 \lambda \right)}{\beta_{0}^4 (1-2 \lambda)^2}
  +\frac{(1-2 \lambda) \ln (1-2 \lambda ) \left(\beta_{0} \beta_{2} 
    (1-2 \lambda )+2 \beta_{1}^2 \lambda \right)}{2\beta_{0}^4 (1-2 \lambda)^2} 
  \notag\\
  &\hspace{10mm}+\frac{\beta_{1}^2}{4 \beta_{0}^4}
  \frac{(1-4 \lambda ) \ln ^2(1-2 \lambda )}{(1-2 \lambda)^2}-\frac{\lambda ^2 }{(1-2 \lambda
    )^2} \ln ^2\frac{\mu_R^2}{x_Q^2 \mh^2}\notag\\
  &
  \hspace{10mm}   -\frac{\beta_{1}}{2 \beta_{0}^{2}}\frac{(2 \lambda  (1-2 \lambda)+(1-4 \lambda) \ln (1-2 \lambda ))
  }{(1-2\lambda )^2}\ln\frac{\mu_R^2}{x_Q^2 \mh^2}\bigg)\,.
\end{align}
The expressions of $\tilde{R}'$, $\delta \tilde{R}'$, and
$\tilde{R}''$ used in Eq.~\eqref{eq:NNLL} are defined as
\begin{align}
\tilde{R}' =- \frac{d }{d\tilde L}\left(\tilde{L} g_{1}(\lambda)\right),
~~~~
\delta \tilde{R}' = -\frac{d g_{2}(\lambda)}{d\tilde L},
~~~~
\tilde{R}'' = \frac{d \tilde{R}' }{d\tilde L}.
\end{align}
The $\beta$ function coefficients read
\begin{eqnarray}
  \beta_0 &=& \frac{11 C_A - 2 n_f}{12\pi}\,,\qquad 
  \beta_1 = \frac{17 C_A^2 - 5 C_A n_f - 3 C_F n_f}{24\pi^2}\,,\\
  \beta_2 &=& \frac{2857 C_A^3+ (54 C_F^2 -615C_F C_A -1415 C_A^2)n_f
       +(66 C_F +79 C_A) n_f^2}{3456\pi^3}\,.
\end{eqnarray}
Finally, we have
\begin{align}
  A^{(1)} =& \,2 C_A,
  \notag\\
  \vspace{1.5mm}
  A^{(2)} =&
  \left( \frac{67}{9}-\frac{\pi ^2}{3} \right) C_A^2
  -\frac{10}{9} C_A n_f,
  \notag\\
  \vspace{1.5mm}
  A^{(3)} =&
   \left( -22 \zeta_3 - \frac{67 \pi^2}{27}+\frac{11 \pi^4}{90}+\frac{15503}{324} \right) C_A^3
  + \left( \frac{10 \pi^2}{27}-\frac{2051}{162} \right) C_A^2 n_f\notag\\
  &+ \left( 4 \zeta_3-\frac{55}{12} \right) C_A C_F n_f
  + \frac{50}{81} C_A n_f^2,\notag\\
  \vspace{1.5mm}
  B^{(1)} =&
  -\frac{11}{3} C_A + \frac{2}{3}n_f,
  \notag\\
  \vspace{1.5mm}
  B^{(2)} =&
  \left( \frac{11 \zeta _2}{6}-6 \zeta _3-\frac{16}{3} \right) C_A^2
  + \left( \frac{4}{3}-\frac{\zeta _2}{3} \right) C_A n_f
  + C_A C_F.
\end{align}

\subsection{Matching to fixed order}
\label{sec:matching}

In this section we discuss the matching of the resummed
and the fixed-order results. We work at the level of the
cumulative distribution $\Sigma$, that at NNLO reads
\begin{equation}
\Sigma^{\rm NNLO}(\pt) = \sigma_{\rm tot}^{\rm NNLO} - 
\int_{\pt}^{\infty} {\rm d} \pt' \left[\frac{{\rm d}\sigma}{{\rm d}\pt}\right]^{\rm NLO}.
\end{equation}
We stress that in the main text we only show results for the
differential $\pt$ distribution, therefore we label them as NLO. This
corresponds to what we label as NNLO at the integrated level in this
appendix. Since the total $gg\to H$ cross section is not known in the
full SM beyond NLO, we approximate the NNLO correction to
$\sigma_{\rm tot}^{\rm NNLO}$ by multiplying the exact NLO result by
the NNLO/NLO $K$ factor as computed in the $m_t \to \infty, m_b\to 0$
limit.  We stress, however, that at the level of the differential
distributions we are interested in, this approximation is formally a
N$^3$LL effect, and it is beyond the accuracy considered in our study.

In order to assess the uncertainty associated with the matching
procedure, we consider here two different matching schemes. The first
scheme we introduce is the common additive scheme discussed in the
main text defined as
\begin{equation}
\label{eq:additive}
\Sigma_{\rm add}(\pt) = \Sigma^{\rm NNLL}(\pt) + \Sigma^{\rm NNLO}(\pt) - {\cal T}^{\rm NNLO}\left[\Sigma^{\rm NNLL}(\pt)\right].
\end{equation}
Since the ${\cal O}(\alpha_s^2)$ (relative to the Born) collinear
coefficient functions and virtual corrections are unknown in the full
SM, in the additive scheme we approximate them by multiplying the HEFT
ones by the exact Born squared amplitude. 

The second scheme we consider belongs to the class of multiplicative
schemes.  In the text, we schematically defined it as
\begin{equation}
\label{eq:multiplicative0}
\Sigma_{\rm mult}(\pt) = \Sigma^{\rm NNLL}(\pt) \;{\cal T}^{\rm NNLO}\left[\frac{\Sigma^{\rm NNLO}(\pt)}{\Sigma^{\rm NNLL}(\pt)}\right].
\end{equation}
We recall that we indicate with $\mathcal T^{\rm NNLO}[f]$ the fixed-order expansion of $f$ to NNLO.
The two 
schemes~\eqref{eq:additive},~\eqref{eq:multiplicative0} are equivalent
at the perturbative order we are working at, as they only 
differ by N$^3$LO and N$^3$LL terms. The main difference between the
two schemes is that, in the multiplicative approach, unlike in the
additive one, higher-order corrections are damped by the resummation
factor $\Sigma^{\rm NNLL}$ at low $\pt$. One advantage of the
multiplicative solution is that the NNLO constant terms, of formal
accuracy N$^3$LL, are automatically extracted from the fixed order in
the procedure. Furthermore, as we explained in the text, in this case
higher order effects introduced by the resummation follow the same
scaling in $\pt$ of the fixed-order result, which at least partially
mimics higher order form-factor effects.

However, there is a drawback in using
Eq.~\eqref{eq:multiplicative0} as is. Indeed, $\Sigma^{\rm NNLL}$ does
not tend to one for $\pt\gg Q$, but rather to the luminosity factor
defined in Eq.~\eqref{eq:luminosity-NNLL0} evaluated at
$\tilde{L}=0$. Therefore, the fixed-order result $\Sigma^{\rm NNLO}$
at large $\pt$ receives a relative spurious correction of order
$\alpha_s^3$
\begin{equation}
\Sigma_{\rm mult}(\pt) \sim \Sigma^{\rm NNLO}(\pt)\left(1+{\cal O}(\alpha_s^3)\right).
\end{equation}
Despite being formally of higher order, these effects can be
moderately sizable in processes with large $K$ factors such as Higgs
production. There are different possible solutions to this problem. In
Ref.~\cite{Bizon:2017rah} the resummed factor (and the relative
expansion) was modified by introducing a damping factor as
\begin{equation}
\Sigma^{\rm NNLL} \to \left(\Sigma^{\rm NNLL}\right)^Z,
\end{equation}
where $Z$ is a $\pt$-dependent exponent that effectively acts as a
smoothened $\Theta$ function that tends to zero at large $\pt$. This
solution, however, introduces new parameters that control the scaling
of the damping factor $Z$ (see Section 4.2 of
Ref.~\cite{Bizon:2017rah} for details). In this article we adopt a
simpler solution, which avoids the introduction of extra parameters in
the matching scheme. We therefore define the multiplicative matching
scheme by normalizing the resummed prefactor to its asymptotic value
for at $\tilde{L}\to 0$. This is simply given by 
\begin{equation}
\label{eq:asypt}
\Sigma^{\rm NNLL}_{\rm asym.} = \lim_{\tilde{L}\to 0}
{\cal L}(k_{\perp 1}).
\end{equation}

We obtain
\begin{equation}
\label{eq:multiplicative1}
\Sigma_{\rm mult}(\pt) = \frac{\Sigma^{\rm NNLL}(\pt)}{\Sigma^{\rm NNLL}_{\rm asym.} } {\cal T}^{\rm NNLO}\left[\Sigma^{\rm NNLL}_{\rm asym.} \frac{\Sigma^{\rm NNLO}(\pt)}{\Sigma^{\rm NNLL}(\pt)}\right],
\end{equation}
where
\begin{equation}
\Sigma^{\rm NNLL}(\pt)\xrightarrow[\pt \gg Q]{} \Sigma^{\rm NNLL}_{\rm asym.} .
\end{equation}
This ensures that in the $\pt \gg Q$ limit
Eq.~\eqref{eq:multiplicative1} reproduces by construction the
fixed-order result, and no large spurious, higher-order, corrections
arise in this region. The detailed matching formulas for the two
schemes considered in our analysis are reported below.

We start by introducing a convenient notation for the
perturbative expansion of the various ingredients. We define
\begin{align}
\sigma_{\rm tot}^{\rm NNLO}  = \sum_{i=0}^2\sigma^{(i)},\qquad
\Sigma^{\rm NNLO}(\pt) = \sigma^{(0)} + \sum_{i=1}^2\Sigma^{(i)}(\pt),
\end{align}
where 
\begin{align}
\Sigma^{(i)}(\pt) = \sigma^{(i)} + \bar{\Sigma}^{(i)}(\pt),\qquad \bar{\Sigma}^{(i)}(\pt) \equiv - \int_{\pt}^{\infty} {\rm d}{\pt'} \frac{{\rm d} \Sigma^{(i)}({\pt'})}{{\rm d} {\pt'}}.
\end{align}
Moreover, we denote the perturbative expansion of the resummed cross
section $\Sigma^{\rm
  NNLL}$ as
\begin{equation}
{\cal T}^{\rm NNLO}\left[\Sigma^{\rm NNLL}(\pt)\right] = \sigma^{(0)} + \sum_{i=1}^2 \Sigma_{\rm NNLL}^{(i)}(\pt).
\end{equation}
With this notation, the additive scheme of Eq.~\eqref{eq:additive}
becomes (for simplicity we drop the explicit dependence on $\pt$ in the following)
\begin{align}
          \Sigma_{\rm add} =& \Sigma^{\rm NNLL}+ \left\{\sigma^{(1)}+\bar \Sigma^{(1)}- \Sigma_{\rm NNLL}^{(1)}\right\}+  \left\{\sigma^{(2)}+\bar  \Sigma^{(2)}
                - \Sigma_{\rm NNLL}^{(2)}\right\},
\end{align}
where the three terms in curly brackets denote the NLO, NNLO and
N$^3$LO contributions to the matching, respectively.

For the multiplicative scheme we need to introduce the perturbative
expansion of the asymptotic value $\Sigma^{\rm NNLL}_{\rm asym.}$, defined in
Eq.~\eqref{eq:asypt}. We write
\begin{equation}
\Sigma^{\rm
  NNLL}_{\rm asym.} = \sigma^{(0)} + \Sigma_{\rm asym.}^{(1)}.
\end{equation}
With this notation the matching formula~\eqref{eq:multiplicative1} reads
\begin{align}
&	\Sigma_{\rm mult}(\pt) =  \frac{\Sigma^{\rm NNLL}}{\Sigma^{\rm
  NNLL}_{\rm asym.}}\Bigg[\sigma^{(0)}+ \left\{\sigma^{(1)}+\bar
  \Sigma^{(1)} + \Sigma_{\rm asym.}^{(1)}- \Sigma_{\rm NNLL}^{(1)} \right\}\nonumber 
	+ \Bigg\{\sigma^{(2)} + \bar \Sigma^{(2)} 
\\
&
- \Sigma_{\rm NNLL}^{(2)}
 +\frac{\Sigma_{\rm asym.}^{(1)}}{\sigma^{(0)}}(\sigma^{(1)}+\bar \Sigma^{(1)}) 
       + \frac{(\Sigma_{\rm NNLL}^{(1)})^2}{\sigma^{(0)}} - \frac{\Sigma_{\rm NNLL}^{(1)}}{\sigma^{(0)}}(\sigma^{(1)}+\bar \Sigma^{(1)}+\Sigma_{\rm asym.}^{(1)}) \Bigg\}\Bigg],
\end{align}
where, as above, we grouped the terms entering at NLO, and NNLO within curly brackets.

\clearpage

\bibliographystyle{JHEP}
\bibliography{hjet}

\end{document}